\documentclass[aip,amsmath,amssymb,reprint]{revtex4-2}
\pdfoutput=1

\usepackage[colorlinks,linkcolor=blue,citecolor=blue,urlcolor=blue]{hyperref} 

\usepackage{graphicx}
\usepackage{dcolumn}
\usepackage{bm}
\usepackage{soul}

\usepackage[utf8]{inputenc}
\usepackage[T1]{fontenc}

\begin{document}

\title{Single-cell Bayesian deconvolution}

\author{Gabriel Torregrosa}
\affiliation{Department of Medicine and Life Sciences, Universitat Pompeu Fabra, Barcelona Biomedical Research Park, 08003 Barcelona, Spain}

\author{David Oriola}
\affiliation{EMBL Barcelona, Dr. Aiguader 88, 08003 Barcelona, Spain}

\author{Vikas Trivedi}
\affiliation{EMBL Barcelona, Dr. Aiguader 88, 08003 Barcelona, Spain}
\affiliation{EMBL Heidelberg, Developmental Biology Unit, 69117 Heidelberg, Germany}

\author{Jordi Garcia-Ojalvo}
\affiliation{Department of Medicine and Life Sciences, Universitat Pompeu Fabra, Barcelona Biomedical Research Park, 08003 Barcelona, Spain}

\begin{abstract}
Individual cells exhibit substantial heterogeneity in protein abundance and activity, which is frequently reflected in broad distributions of fluorescently labeled reporters.
Since all cellular components are intrinsically fluorescent to some extent, the observed distributions contain background noise that masks the natural heterogeneity of cellular populations.
This limits our ability to characterize cell-fate decision processes that are key for development, immune response, tissue homeostasis, and many other biological functions.
It is therefore important to separate the contributions from signal and noise in single-cell measurements.
Addressing this issue rigorously requires deconvolving the noise distribution from the signal, but approaches in that direction are still limited.
Here we present a non-parametric Bayesian formalism that performs such a deconvolution efficiently on multidimensional measurements, in a way that allows estimating confidence intervals precisely.
We use the approach to study the expression of the mesodermal transcription factor Brachyury in mouse embryonic stem cells undergoing differentiation. 
\end{abstract}

\maketitle

\section*{Introduction}

The inherent stochasticity of biological processes leads to substantial heterogeneity even among genetically identical cells in the same environment \citep{Elowitz:2002wq,suel2007,Chang:2008th}.
The degree to which this heterogeneity affects, or even dictates, cellular decision-making in most situations is still an open question.
This issue is of paramount importance in processes such as mammalian development, where hundreds (if not thousands \citep{osumisutherland2021cell}) of distinct cell types (cell states) emerge from a small number of identical undifferentiated cells \citep{Nowotschin:2019uz,Weinreb:2020wj,Mittnenzweig:2021vt}.
Identifying the molecular mechanisms underlying these cell-fate decision programs and their interplay with cellular heterogeneity \cite{Garcia-Ojalvo:2012vd,Rue:2013uy} requires a rigorous quantification of cellular states across large numbers of cells.

Flow cytometry enables monitoring the distributions of abundances and activities of selected proteins for thousands of cells at a time, using fluorescently labeled markers.
Cells, however, have a non-negligible amount of autofluorescence in the emission spectrum of most fluorescent probes  (500 to 700 nm).
This leads to a background noise that must be subtracted from the total signal emitted by fluorescently labeled cells \citep{Corsetti:1988tj}, in order to adequately relate the signal distribution provided by the cytometer to the mechanisms regulating the expression and/or activity of the protein of interest.
Several standard methods exist for addressing this issue on a cell-by-cell basis. One can use, for instance, additional fluorescence channels outside the emission spectrum of the fluorophores serving as regressor variables \citep{Roederer:1986uk,Alberti:1987ud}. Alternatively, a second laser system can provide an independent measurement of both signal and autofluorescence \citep{Steinkamp:1986ts}.
Beyond issues of cost or accessibility, the effectivity of such solutions is limited, because there is no guarantee that autofluorescence from another channel, or from another excitation source, is a good proxy for autofluorescence in our channel of interest.

Commonly, rather than dedicating measurement resources to assess autofluorescence, control measurements of unlabeled cells are used to set a baseline of the signal coming from the naturally present autofluorescent components in the cell.
This procedure, however, does not lead to a quantitative determination of the distribution of the signal coming \emph{exclusively} from the fluorescent probe.
Such quantitative assessment would require deconvolving the fluorescence distribution obtained in labeled cells from the one produced by unlabeled cells.
Here, we propose a non-parametric Bayesian approach to this deconvolution problem, applicable to multichannel measurements.
The method is robust and efficient, requiring cell numbers not larger than those typically considered in standard flow cytometry runs, and gives natural confidence intervals of the target distributions, which makes it attractive for a variety of applications.

There is an extensive statistics literature addressing the additive deconvolution problem\footnote{We do not discuss in what follows other deconvolution problems that are not relevant to our situation, such as those related with composition of distributions, which usually require complete knowledge of the noise distribution \citep{Efron:2016td,GDAlternative}.}.
A common set of deconvolution methods are kernel-based approaches, such as those relying on Fourier transforms \citep{Stefanski:1990uw,Liu:1990vw,Fan:1991wz,Diggle:1993tv,FFT,Comte:2011uq,Delaigle:2014te}, which use the fact that in Fourier space, a deconvolution is simply the product of two functions.
Two problems arise from such methods that limit their applicability in practical cases.
First, Fourier transforms (and other methods that use orthogonal local bases, such as wavelets \citep{Fan:2002wm}) are not positive defined.
Consequently, kernel-based methods lead to deconvolved pseudo-distributions with artificial features, which are hardly interpretable for practical applications.
Second, these methods usually lead to point estimates, and therefore do not provide native confidence intervals (i.e. without applying additional statistical approximations) that allow us to assess the quality of the inferred target distribution.

A second class of deconvolution approaches are likelihood-based methods \citep{Lee:2013tm}, which estimate the unknown target distribution using maximum likelihood approaches. As in the case of kernel-based methods, these approaches provide us with point estimates, and usually assume exact knowledge of the noise distribution.
Finally, a third class of methods involve Bayesian inference \citep{Staudenmayer:2008uu,GDHeterogeneity,GDHeterogeneity2}, which does not require complete knowledge of the noise distribution and naturally provides confidence intervals of the estimates obtained.
So far, however, these Bayesian methods have been applied to repeated measurements of the same individual entities that are being monitored (in our case, cells), which is not a realistic possibility in standard flow cytometry.
Our non-parametric Bayesian approach does not require repeated measurements and retains all the above-mentioned advantages of Bayesian methods.
We have implemented the procedure in a Julia package available in GitHub (\href{https://github.com/dsb-lab/scBayesDeconv.jl}{https://github.com/dsb-lab/scBayesDeconv.jl}). 

The work is structured as follows. First, we introduce the mathematical description of the additive convolution problem in the context of flow cytometry data, and explain our proposed deconvolution method based on non-parametric Bayesian models.
Second, we validate our method using synthetic datasets with known target distributions, and compare its results to other existing methods.
Third, we further test our method in real flow-cytometry data of mouse embryonic stem cells undergoing differentiation. We test the dataset in two conditions. Initially, we treat our cells with a low concentration of a fluorescent dye (which masks the real flow-cytometry signal and acts as an external noise), and validate our method by deconvolving the noise coming from the dye and comparing it to the control case where the dye was not added. This offers evidence of the robustness of the method in real applications while having a ground truth. Additionally, we perform the deconvolution using several channels from the dataset, to show the effectiveness of the method in multichannel measurements, and comment on the complementarity of our approach to other processing steps common in analysis pipelines of flow cytometry data. 
We conclude by discussing the limitations of the method and possible ways to improve it in future work.  


\section*{Results}

\subsection*{Theoretical definition of the problem}

Consider a population of cells containing fluorescent markers that label the abundances or activities (e.g. phosphorylation states) of certain proteins of interest.
Flow cytometry measurements provide us with the distribution $p_{c}(C)$ of total fluorescence signal $C$ emitted by each individual cell in the population, as measured by the flow cytometer detectors.
These signals have two components: the fluorescence $T$ emitted exclusively by the target fluorophores that report on the proteins of interest, and the autofluorescence $\xi$ emitted by cellular components \emph{other} than our fluorescent labels:
\begin{equation}
C=T+\xi
\label{eq:problem}
\end{equation} 
If these two components are independent of one another, the distribution $p_c(C)$ of the measured signal takes the form of a \emph{convolution} of the distributions of $T$ and $\xi$:
\begin{equation}
(p_T*p_\xi)(C) := \int_{0}^\infty \!\!p_T(C-\xi)\,p_\xi(\xi)\,d\xi 
\label{eq:conv}
\end{equation}
Similarly to $p_c$, the distribution $p_\xi$ of the autofluorescence $\xi$ can be measured in the flow cytometer by using unlabeled cells that are otherwise identical to the labeled ones.
On the other hand, the probability distribution of $T$, $p_T$, cannot be measured directly.
Our goal is to extract (deconvolve) the distribution $p_T$ from the measured distributions $p_c$ and $p_\xi$, considering 
that we only have a finite set of samples (cells) of $C$ and $\xi$. 

In what follows, we first introduce the way in which we describe the distributions involved in the problem.
Next we define the posterior distribution given by the model and the data, and finally we discuss the methods used to explore the parameter space for the deconvolution problem.

\paragraph*{Mixture model.}
The vast majority of real datasets (including those generated in flow cytometry) result from complex combinations of variables that cannot be explained in general with simple distributions.
In order to adapt flexibly to such conditions, we use mixtures of probability distributions as our basis set.
Any 
function can be arbitrarily well approximated using an adequate choice of basis functions, provided enough components are included in the mixture.
We can describe both the target and the noise distributions by independent mixtures of $K$ components:
\begin{equation}
p(x|\bm{\phi}) = \sum_{\eta=1}^K \omega_\eta\mathcal{B}(x|\bm{\psi}_\eta)
\label{eq:mixture}
\end{equation}
where $x$ is multivariate in the case of multichannel measurements. $\omega_\eta$ denotes the weight of each base $\mathcal{B}(x|\bm{\psi}_\eta)$ in the mixture, and $\bm{\psi}_\eta$ represents its parameters (the means and covariance matrices in the case of multivariate normal distributions)\footnote{Throughout the paper we use greek subindices, in particular $\eta$ and $\lambda$, when summing over basis functions, and roman subindices when summing over samples --cells--.}.
The sets of $\omega_\eta$ and $\bm{\psi}_\eta$ are in turn represented by the vector $\bm{\phi}$ in what follows.
Using these basis functions, the distribution of the observed signal $C$ can be described as a superposition of all the convolutions between them:
\begin{align}
p_c(c|\bm{\phi}_T,\bm{\phi}_\xi)&= \sum_{\eta=1}^{K_T}\sum_{\lambda=1}^{K_\xi} \omega^T_\eta\omega^\xi_\lambda(\mathcal{B}\ast\mathcal{B})(c|\bm{\psi}^T_\eta,\bm{\psi}^\xi_\lambda)
\label{eq:convolution}
\end{align}
where $(\mathcal{B}\ast\mathcal{B})(c|\psi^T_\eta,\psi^\xi_\lambda)$ represents the convolution of two basis distributions with parameters $\bm{\psi}^T_\eta$ and $\bm{\psi}^\xi_\lambda$, respectively, and $K_T$ and $K_\xi$ denote the number of bases used in each of the two mixtures. 

The choice of basis functions to describe our data is crucial.
We propose to use multivariate normal distributions to describe the data.
Normal distributions with unknown mean and covariance have been shown to be flexible enough to represent datasets with high quality, requiring less components than more rigid methods. They have the additional advantage that its convolution turns out to be another normal distribution with modified parameters \citep{Gelman}. This last property is especially interesting for deriving exact analytic results for the Bayesian sampling process.

\paragraph*{Posterior distribution and likelihood.}

Our goal is to extract, starting from samples of the distributions of total signal $C$ and noise $\xi$, the parameters of the distribution of the target signal $T$.
Defining the problem in terms of probability distributions leads very naturally to work with Bayesian methods.
According to Bayes' rule, the posterior distribution that represents the probability of the parameters given the data is
\begin{equation}
p(\bm{\phi}_T,\bm{\phi}_\xi|\mathbf{c},{\boldsymbol\xi})\propto \Big[p(\mathbf{c}|\bm{\phi}_T,\bm{\phi}_\xi)p(\bm\xi|\bm{\phi}_\xi)\Big]
\Big[p(\bm{\phi}_T)p(\bm{\phi}_\xi)\Big],
\label{eq:b_rule}
\end{equation}
where $\mathbf{c}=\{c_i:i=1\ldots N_c\}$ and $\bm\xi=\{\xi_i:i=1\ldots N_\xi\}$ are the sets of observed samples of the total signal and the noise, respectively, with $N_c$ and $N_\xi$ representing the number of samples in each case.

The first bracket on the right-hand side of Eq.~(\ref{eq:b_rule}) is the likelihood function, which corresponds to the probability of the data given the parameters. Under the assumption of independent and identically distributed (iid) observations, this function is given by
\begin{equation}
\mathcal{L}(\bm{\phi}_T,\bm{\phi}_\xi) = \prod_{i=1}^{N_c} p_c(c_i|\bm{\phi}_T,\bm{\phi}_\xi)\prod_{j=1}^{N_\xi} p_\xi(\xi_j|\bm{\phi}_\xi)
\label{eq:likelihood}
\end{equation}
As can be seen, the information about the noise parameters is contained in both datasets, while the information about the target parameters only appears in the convolved data. 

As shown in Eq.~\eqref{eq:b_rule} above, the posterior distribution can be estimated by multiplying the likelihood by the prior distribution of parameter values.
Since the datasets required for a good deconvolution are large, the impact of the prior distribution should be negligible.
Therefore, the posterior landscape is effectively described by \eqref{eq:likelihood}, and the prior distribution is only present in our approach for formal and computational reasons.
In any case, since no prior information exists on the parameters that describe the noise and target mixture components, we impose vague priors over the plausible set of parameters (see Supplementary Sec.~S4). 

\paragraph*{Sampling.}

The posterior distribution is a complex multimodal (multi-peaked) object containing all the configurations of the target and noise distributions that are consistent with the observed data.
A correct and efficient exploration of this distribution is essential to set bounds on the candidate models that explain the data. The Bayesian model, as stated in Eq.~\eqref{eq:b_rule}, has no direct closed form. Sampling from this model as stated would require the introduction of complex sampling algorithms such as Markov Chain Monte Carlo (MCMC) \citep{Gelman} and nested sampling \citep{NS_main}. These sampling processes are computationally very expensive and scale linearly with the sample size, which in the case of flow cytometry datasets is of the orders of tens to hundreds of thousands samples per dataset.

In order to overcome these shortcomings, we propose two approximations to the posterior distribution. The first one decouples the inference of the noise and target distributions parameters as two independent problems (see Supplementary Section S2.1). The parameters of the noise distribution can be fitted using well-known mixtures of normal distributions, with the target being described by a convolved mixture model. The second approximation simplifies the sampling of the convolved mixture model to enable sampling using standard Gibbs sampling algorithms (see Supplementary Section S4.2). The combination of both approximations allows the posterior to be sampled using Gibbs sampling, which is an efficient MCMC method of sampling from the posterior.

In addition to these approximations, we offer two alternative Bayesian mixtures, namely finite or infinite normal mixture models \cite{li_tutorial_2019}. The latter approach has the added advantage over the former of not requiring fixing the number of basis functions to describe the data. However, for most practical purposes, finite normal mixture models are able to fit the data with high accuracy and in a more efficient manner regarding how the corresponding algorithms work.

\begin{figure*}[htb]
\centerline{\includegraphics[width=0.8\textwidth]{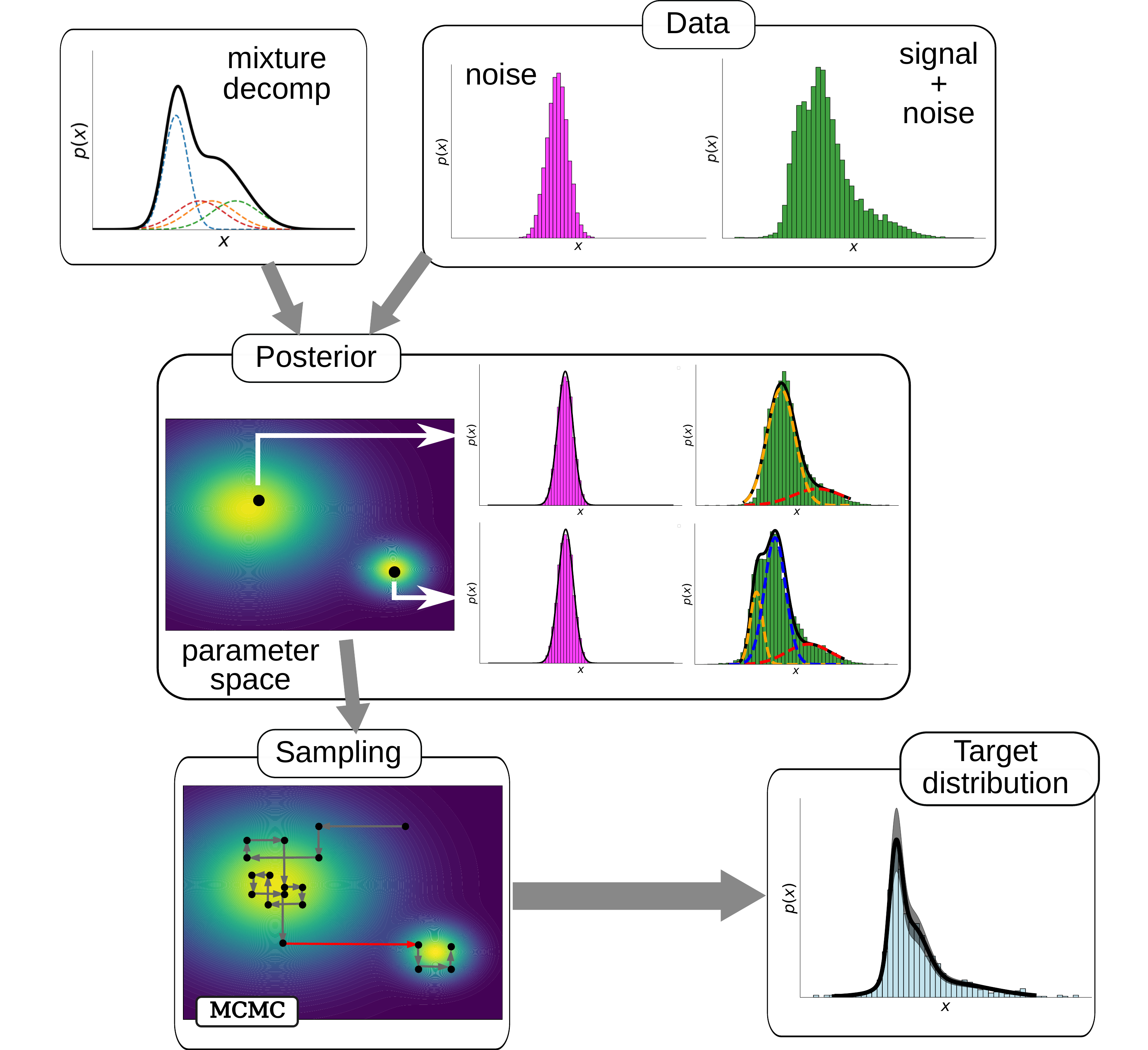}}
\caption{Scheme of the deconvolution process.
The signal and noise mixture distributions, together with the observed data (top row), define the posterior distribution over the parameter space of the mixture, Eq.~\eqref{eq:likelihood} (middle row). This distribution can present multiple peaks, sometimes degenerate with respect to basis label exchange, each corresponding to a different mixture description of the observed and target distributions. The red arrow in the Gibbs Markov Chain Monte Carlo sampling plot (bottom left) represents an unlikely jump between two peaks separated by a relatively wide probability valley.
}
\label{fig:01}
\end{figure*}

\paragraph*{Analysis pipeline.}

Given the concepts and tools described above, the analysis of the data is performed as follows (Fig.~\ref{fig:01}).
First, the distributions of the target and autofluorescence signals are assumed to be described by mixtures (top left panel in Fig.~\ref{fig:01}), as defined by Eqs.~(\ref{eq:mixture}) and (\ref{eq:convolution}).
The data measured experimentally correspond to the autofluorescence signal (noise) and the total signal of the labeled cells (which includes the autofluorescence), as shown in the top right panel of Fig.~\ref{fig:01}.
The mixture assumption and the observed data samples allow us to construct the posterior distribution (Eq.~\eqref{eq:b_rule} and middle panel in Fig.~\ref{fig:01}) from the likelihood function defined by Eq.~(\ref{eq:likelihood}) and the prior distributions discussed in the Supplementary Sec.~S4. The posterior distribution is a function of the model parameters (weights of the mixtures and parameters of the basis functions). We next explore (sample) the posterior distribution in parameter space using the Gibbs sampling approximation (bottom left panels in Fig.~\ref{fig:01}). 
This algorithm provides us with a representative sampling of the parameters of the target distribution, which allows us to compute an average of this distribution and its confidence interval (bottom right panel in the figure).

\subsection*{Application to synthetic data}

\begin{figure*}[htbp]
\includegraphics[width=\textwidth]{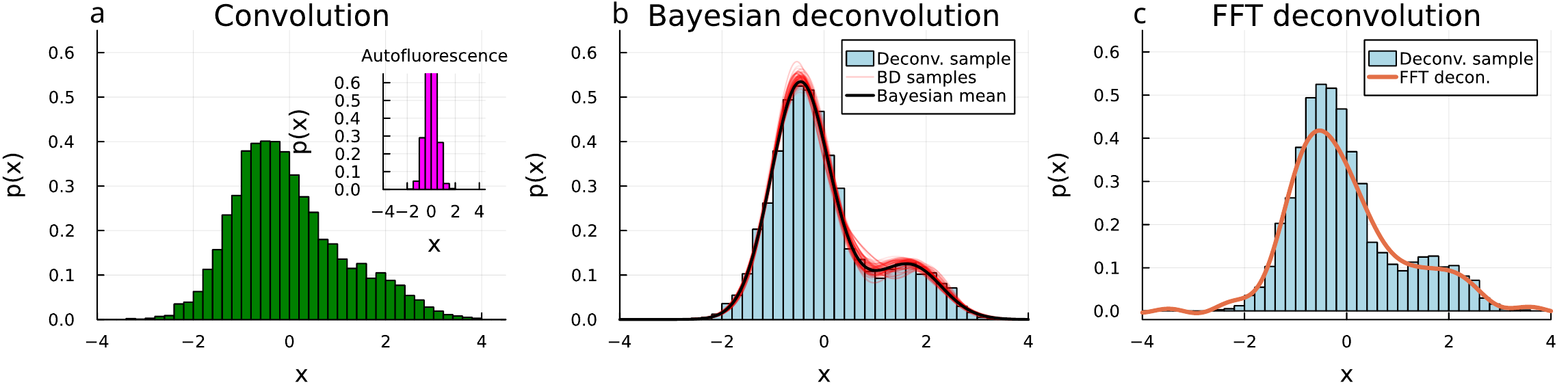}
\caption{Deconvolution instance for a target bimodal distribution (light blue in panels b and c) corrupted by a normally distributed noise with a $\text{SNS}=2$ (magenta distribution in the inset of panel a), applying the Bayesian Deconvolution method (panel b) and the FFT method (panel c).
The total distribution from which the noise is deconvolved is shown in green in panel (a).
The red lines in panel (b) depict samples of the Bayes fitting process.
In this case, the noise distribution is a single normal function with mean $\mu^\xi=0$ and standard deviation $\sigma^\xi=0.5$, and the target is a mixture of two normal functions with means $\mu_1^T=-0.43$ and $\mu_2^T=1.67$, standard deviations $\sigma_1^\xi=\sigma_2^\xi=0.6$, and weights $\omega_1^T=0.8$ and $\omega_2^T=0.2$.}
\label{fig:02}
\end{figure*}

First, we benchmark the ability of our method to recover the target distribution by using collections of synthetic datasets.
In those datasets, the target and noise distributions are known, so we can compare the result of the deconvolution against a ground truth.
To do this, we applied our method to the synthetic data described in Sec.~\ref{sec:synthetic}.
To prove the robustness of the implementation, we ran the test using four components, both for the noise and target distributions.
We also avoided checking the full convergence of the algorithm manually.
We ran the algorithm in this suboptimal conditions to avoid having to fine tune the specific parameters, which could lead to positive bias favoring our method in comparison with FFT-based approaches.
We contrasted the results of our method with those of a specific FFT approach that does not require knowledge of the autofluorescence distribution \cite{FFT}.
Figure \ref{fig:02} shows a typical instance of the deconvolution performance of the two methods.
Panel (a) shows in green the total (convolved) signal mimicking the output of a flow cytometry experiment.
In this synthetic case, the signal is obtained by forward convolving a target distribution with the characteristics given above, shown in light blue in panels (b) and (c), with a noise (auto-fluorescence) distribution, shown in magenta in the inset of panel (a).
The goal in this case is to recover (deconvolve) the ground-truth target distribution (panels b and c in Fig. \ref{fig:02}) from the total and noise distributions (panel a).
Figures \ref{fig:02}(b,c) show that our Bayesian deconvolution method recovers the target distribution well. In comparison, the FFT method leads to oscillatory components in the deconvolution, and correspondingly to artefactual negative values in the probability distribution.
Additionally, the Bayesian deconvolution method naturally provides a confidence interval, shown by the different samples (red lines) in panel (b) of Fig. \ref{fig:02}.

For benchmarking purposes, we compared the deconvolved distribution with the real target distribution, which is known in this case, using the mean integrated overlap (MIO), as defined in Supplementary Sec.~S6.
We preferred this measure to the mean integrated squared error (MISE), which is commonly used in the deconvolution literature for theoretical reasons \cite{Comte:2011uq,FFT,Liu:1990vw,Fan:2002wm}, since the MIO measure is easier to interpret, as it corresponds directly to the absolute overlap of two probability distributions.
We also avoid more common measures of distribution dissimilarity such as the Kolmogorov-Smirnov test, since such methods would underestimate the ability of FFT-based methods to converge to the ground-truth deconvolution, given that they lead to artifacts in the resulting distributions, as shown above.

\begin{figure}[htbp]
\centerline{\includegraphics[width=0.45\textwidth]{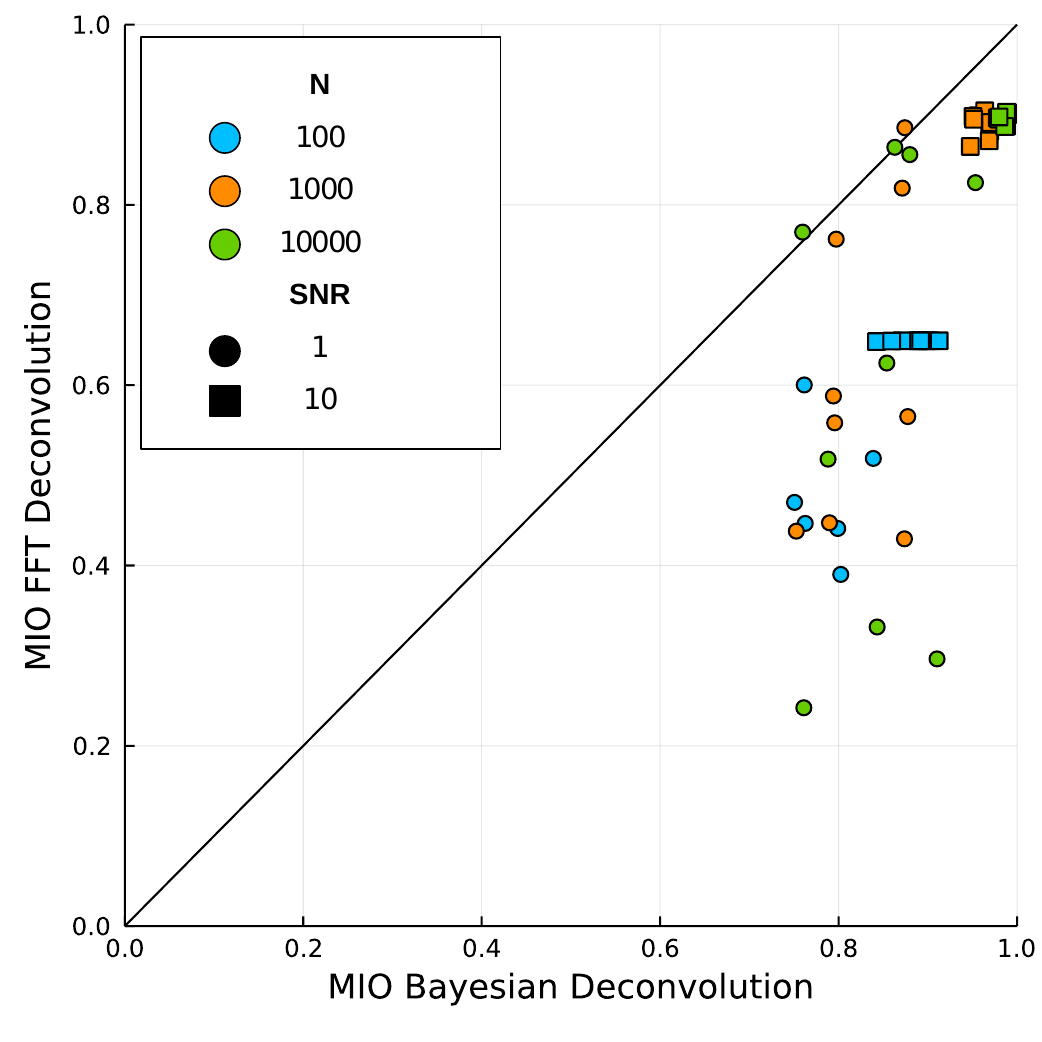}}
\caption{Similarity between the deconvolved and ground-truth target distributions as expressed by the Mean Integrated Overlap (MIO) for the two deconvolution methods (x and y axis), at different sampling sizes for the synthetic datasets described in Sec.~\ref{sec:synthetic}, with $\textrm{SNR}=1$ (circles) and $\textrm{SNR}=10$ (squares).}
\label{fig:03}
\end{figure}

Figure \ref{fig:03} compares the deconvolution efficiency of the FFT-based method and our single-cell Bayesian deconvolution approach, in terms of the MIO measure that quantifies the similarity between the convolved distribution and the real one.
According to its definition (see Supplementary Sec.~S6), a MIO value of 1 corresponds to a perfect overlap, while the measure is 0 when two distributions do not overlap at all. 
As can be seen in Fig.~\ref{fig:03}, the single-cell Bayesian Deconvolution method outperforms the Fourier-based method in almost all the cases considered, the difference being more substantial for high levels of noise (small SNR, circles).
In particular, the overlap is never below 0.7 for the Bayesian methods, while it can reach values near 0 in the FFT case depending on the type of distributions involved, particularly for low sampling numbers.

Additionally, it is worth noting that while the single-cell Bayesian deconvolution method is able to reproduce almost perfectly the target distribution (MIO close to 1) for large enough dataset sizes, the Fourier-based method always saturates to a suboptimal level even as the number of samples increases, even for low noise levels (large SNR, crosses in Fig.~\ref{fig:03}).
Qualitatively, this is due to the oscillatory features in the distribution produced by the noise during the deconvolution with the FFT method, as a consequence of the oscillatory nature of the Fourier basis functions. In general, it has been proved that in FFT methods the convergence to the actual distribution grows sublinearly with the sample size, with scaling dependencies that make the method unfeasible for practical purposes \cite{FFT,GDHeterogeneity,Liu:1990vw}.
Since Bayesian deconvolution makes use of local basis functions, more parsimonious solutions can be obtained, preventing the degradation of the target distribution due to the noise.     

In addition to the direct comparison performed between FFT and Bayesian deconvolutions, we compared the results of the deconvolution against a model that simply fits the convolved data ignoring the convolution problem. A condition to impose on any practical deconvolution algorithm is that the result of the deconvolution is not worse than not doing the decomposition at all. The results of this comparison can be observed in Supplementary Figure S2. The results show that, while the Fourier-based method generally tends to make worse predictions than the null model (fitting without deconvolution), the Bayesian approach consistently generates results virtually as good as when the noise is negligible (SNS=10), and improves systematically the distribution in the high-noise case (SNS=1) over the null model.
In conclusion, the Bayesian method that we propose has combined benefits, both qualitatively and quantitatively, over other basis methods as FFT and consistently improves the result (or at least not degrades the quality of the distribution).

\subsection*{An experimental dataset with external noise}

\begin{figure*}[htbp]
\centerline{\includegraphics[scale=0.25]{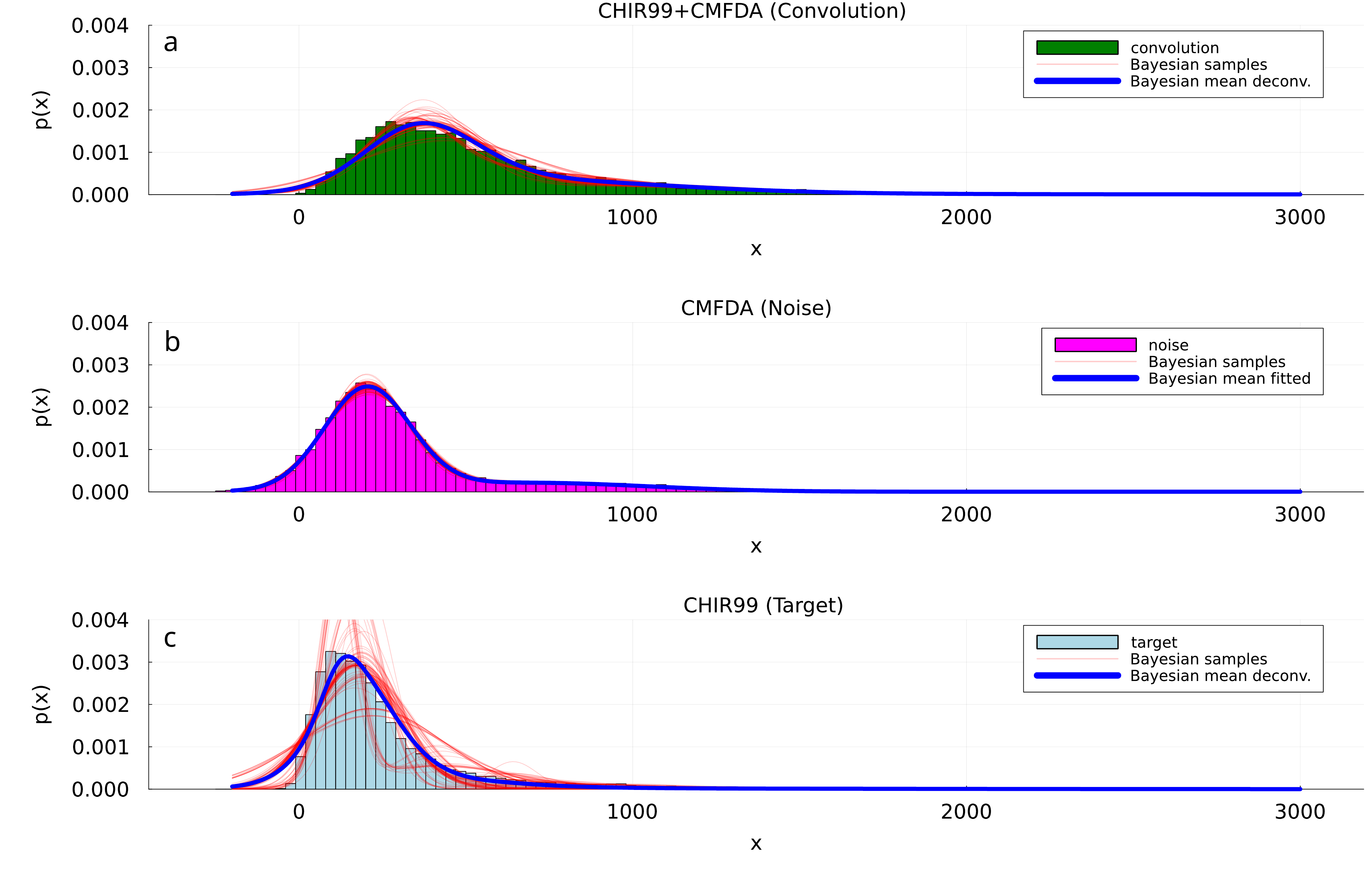}}
\caption{Deconvolution of the CMFDA dye in condition c4 (N2B27+CHIR99+CMFDA). Overlaying the distributions we show realizations of the Bayesian sampling process (red lines) for the three distributions (convolution (a), noise (b) and target (c)) obtained during the fitting. The convolved and target distributions (green and light blue, respectively) come from the real data. The distribution of the dye (in magenta (b)) results from sampling the inferred dye distribution, as described in the text.}
\label{fig:04}
\end{figure*}

Next we applied our method to an experimental dataset in which the noise distribution is unknown. 
To generate a ground truth, we use cells with the Brachyury reporter T/Bra::GFP in two media conditions (N2B27 supplemented with $3$ $\mu$M CHIR99 and DMSO as control). At day 3, for each condition, one of the replicas was treated with 20 nM Green CMFDA dye and incubated for $3$ min prior to flow cytometry, and the second one was incubated for 3 min with N2B27 (see Methods). Given that the dye incorporates in the cell cytoplasm and its emission spectrum is similar to the one of GFP, the dye acts as a noise source to the GFP signal coming from the Brachyury reporter.
Consequently, we have the following four conditions, with their potential outcomes in terms of the signal measured in the FITC-A channel:
\begin{itemize}
\item[(c1)] \textbf{Control}: Brachyury expression is minimal, and the signal comes mainly from the intrinsic autofluorescence of the cells.
\item[(c2)] \textbf{Control$+$CMFDA}: Brachyury expression is minimal, and the signal comes mainly from the CMFDA dye, plus intrinsic autofluorescence of the cells.
\item[(c3)] \textbf{CHIR99}: Brachyury expression is upregulated, and thus the signal contains both the T/Bra::GFP reporter and the intrinsic autofluorescence of the cells.
\item[(c4)] \textbf{CHIR99$+$CMFDA}: Brachyury expression is upregulated, and thus the signal contains both the T/Bra::GFP reporter, the signal from the CMFDA dye and the intrinsic autofluorescence of the cells.
\end{itemize}
The four signal distributions corresponding to these four conditions are shown in Supplementary Fig.~S3. We note that the distribution of the signal coming only from the dye cannot be measured independently, given the intrinsic autofluorescence of the cells.
Moreover, the dye might be absorbed differently depending on the state of the cells (CHIR99 treatment), and the low concentration of the dye ($\sim$ nM) might lead to significant cell-to-cell variabilty on the absorbance of the dye. These features make it a suitable dataset to test the robustness of our method.
We first used our method to deconvolve the signal of the dye from the total signal observed in the experimental conditions (c2), using condition (c1) to define the `noise' distribution.
The results of the deconvolution are shown in Supplementary Fig.~4, which presents the dye distribution deconvolved from experimental conditions (c1) and (c2).
Once the dye distribution was obtained from the deconvolution of conditions (c1) and (c2), we tested the consistency of our approach by considering the dye signal as the noise in condition (c4).
Deconvolving the dye distribution from the one measured in (c4) should lead to the distribution obtained in condition (c3), which we can measure experimentally and thus serves as the ground truth in this case.
The result can be seen in Fig.~\ref{fig:04}.
Even in this case, where we use a deconvolved dataset for a second deconvolution, we observe that the inferred distribution resulting from our method is in good agreement with the experimentally measured distribution in condition (c3) ($\textrm{MIO}=0.81\pm 0.04$) and improves over the null model that ignores the noise ($\textrm{MIO}=0.46\pm 0.01$).

Overall, this experiment shows that our method is robust even when our conditions (target and noise independence) are not fully met, and over other fluctuations present in real datasets.

\subsection*{Multidimensional distributions}

Finally, we applied our method in a real flow cytometry dataset with multichannel detection.
To that end, we studied the expression of the mesodermal gene Brachyury through a GFP reporter (T/Bra::GFP) in mouse embryonic stem cells (see Methods). In this dataset, two populations are expected to appear: a Brachyury negative population of pluripotent cells, and a positive population capturing the cells exiting from pluripotency towards mesodermal fates.
Cells were treated for 24h with $3$ $\mu$M CHIR99 and $25$ ng/mL Activin A to upregulate Brachyury prior to flow cytometry on day 2 (see Methods).
The signal of the GFP reporter was our target signal, and it was acquired through the FITC-A channel. 

\begin{figure}[htpb]
\centerline{
\includegraphics[scale=0.45]{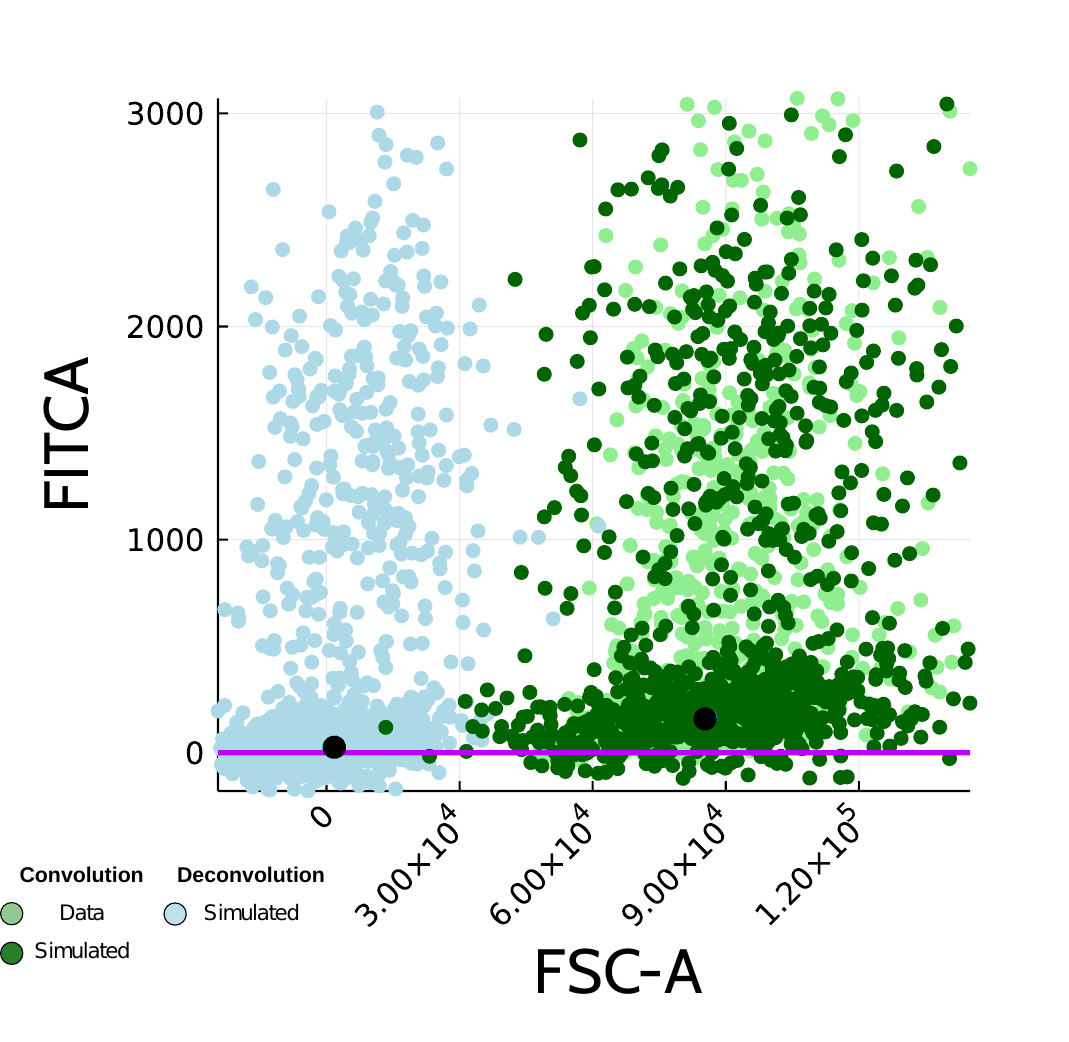}}
\caption{Two-dimensional scatter plot showing a sample of the results of the autofluorescence in a multichannel system. Light and dark green circles are samples of the data from the real data and the fitted convolved distribution, respectively. Light blue circles represent the results of the deconvolution for the two channels. The black dots indicate the mean of the low T:GFP cluster.}
\label{fig:05}
\end{figure}

In a normal flow cytometry dataset, the signal contains the fluorescence emitted by GFP together with the autofluorescence emitted by the cells at the GFP frequency. The effect of the autofluorescence over the T::GFP signal (target) gives the erroneous impression that Brachyury is expressed in the two populations, when looking directly at the convolved data (light green dots in Figure \ref{fig:05}). In other words, it seems there is a non-zero basal expression of the gene in the pluripotent population. 
When deconvolving the data through the Bayes method, however, the result shows that the pluripotent population peaks actually around zero, as expected since these cells are expected not to show any emission of Brachuyry, which is confirmed by the absence of mRNA in the pluripotent states \cite{anlacs2021dynamics}. 

Similar problems of non-negative detection have been reported in other highly autofluorescent populations, as in the case of the false positive detection of MHCII in microglia \cite{li_microglia_2018}, and the false discovery of Foxp3 expressing cell types that are not T cell lineages \cite{li_autofluorescence_2012}.

As can be observed from the other channels (See SSC-A in Figure \ref{fig:05} and all the channels in Figure S5), the deconvolution has an effect in removing the expected variance from all the other channels. The left variance may be an effect of the spillover energy that the GFP fluorophore has in the emission spectrum of other channels as the contribution to broad-band channels as FSC and SSC channels (see for example the effect in the FSC-A in Figure \ref{fig:05}). The use of deconvolved distributions in single fluorophore samples could improve the estimation of spillover coefficients for sample correction \cite{roca_autospill_2021}.

In general, the Bayes deconvolution is straightforwardly applicable to real multichannel datasets and its results are consistent with the expected biological observations, showing a clear correction of defects coming from the presence of autofluorescence in the detection channels.

\section*{Discussion}

In this paper, we propose a Bayesian approach to obtain flow-cytometry distributions of protein abundance or activity convolved with a known source of noise.
The method, which relies on non-parametric Bayesian techniques, is freely available as a Julia package (\href{https://github.com/dsb-lab/scBayesDeconv.jl}{https://github.com/dsb-lab/scBayesDeconv.jl}) and can be used in a straightforward manner in a purely computational way, without the need of dedicated measurement channels or additional laser sources.
It only requires measuring the fluorescently labeled and unlabeled cells and, unlike previously proposed deconvolution methods, it provides well-defined and robust probability distributions, described by mixtures of basis functions.

We measure the quality of the results obtained with our method by comparing the deconvolved distributions with known (ground-truth) target distributions using synthetic data and ad hoc experiments with mouse embryonic stem cells.
We argue that the use of local basis distributions to describe all the distributions involved in the problem (both measured and unknown), and the corresponding use of a relatively small number of degrees of freedom, leads to an efficient inference.
Finally, the Bayesian nature of the method gives rise to a set of candidate target probability distributions.
This reflects the natural indeterminacy present in any process corrupted by noise.
The ability to express indeterminacy in the solutions is crucial for any real application of a deconvolution algorithm. We further show that the method is robust and consistent even under strong noise in real datasets.

The approach that we propose here is applicable to flow cytometry datasets with multiple channels, which is the current outcome of modern flow cytometers. Furthermore, the method is consistent with already well-established methods for flow cytometry processing, such as the calculation of spillover matrices and the reduction of autofluorescence using additional channels for autofluorescence regression, as discussed in the Supplementary text (Section S1.2). In this sense, our deconvolution approach could be applied to calculate corrected one-fluorophore samples for the calculation of the spillover coefficients, or it could be used after calculating the spillover uncoupling with another method, in order to remove the remaining sources of noise in the dataset.

\section*{Limitations of the study}

A potential limitation of the algorithm is the computing time requirements. For $K_n$ mixture components of the noise and $K_t$ components of the target, a single evaluation of the likelihood function scales as $\mathcal{O}(K_n\times K_t \times N)$, where $N$ is the number of measures in the Bayesian model. The current finite mixture implementation is capable of dealing with datasets of $100000$ cells with $K_n=4$ and $K_t=6$ in about 10 minutes in a desktop with an i7 processor. This efficiency will be sufficient for most part of the analysis, but can be a drawback for the analysis of very large datasets when compared with non-Bayesian methods. The computational time required by the infinite mixture model depends strongly on the parameter $\alpha$, which governs the probability of generating new basis functions, but in general this model is more than an order of magnitude slower than its finite counterpart for the same degree of complexity. 
Handling large datasets is a well-established problem in Bayesian statistics, and some lines of work have explored solutions to reduce this computational cost, which might be worth exploring in the context of the single-cell Bayesian deconvolution method proposed here \citep{bigdata1,bigdata2,bigdata3}.

A second potential improvement of the method presented here would be to address the degeneracy under label exchange characteristic of mixture models. Such degeneracy may confound the convergence of the samples to a stationary distribution and the use of the samples for the discovery of clusters, as the same cluster may be assigned to different basis in different samples. A potential solution to this problem would be the implementation of label reassignment methods \cite{cron_efficient_2011}.

\section*{Acknowledgements}

We thank the CRG Tissue Engineering Unit and the UPF/CRG Flow Cytometry Unit for continuous support. This work was supported by the Spanish Ministry of Science and Innovation and FEDER, under projects FIS2017-92551-EXP and PID2021-127311NB-I00, by the ``Maria de Maeztu'' Programme for Units of Excellence in R\&D (grant CEX2018-000792-M), and by the Generalitat de Catalunya (ICREA Academia programme). GT is supported by a FPU doctoral fellowship from the Spanish Ministry of Education and Universities (reference FPU18/05091).
D.O. acknowledges funding from Juan de la Cierva Incorporaci\'on with Project no. IJC2018-035298-I, from the Spanish Ministry of Science, Innovation and Universities (MCIU/AEI).

\section*{Methods}

\subsection*{Synthetic data}
\label{sec:synthetic}

For the target distribution we generated samples from three distributions: symmetric bimodal, asymmetric bimodal and skew symmetric distributions (Supplementary Fig.~S1).
This choice of distributions intends to capture the features present in real datasets, such as the presence of multiple peaks, different cluster sizes, and the generally non-Gaussian character of the data\cite{GDHeterogeneity}. As for the noise, we generated a set of three different noise datasets containing normal, skewed, and Student's t distributions (the last two of which exhibit fat tails) (Supplementary Fig.~S1), in order to test the flexibility of the method against very dissimilar autofluorescence profiles. In order to check the impact of the noise strength, the convolutions between target and noise were generated at two signal-to-noise ratios (SNR).
In our context, we define the SNR as the ratio between target and signal variances for the whole dataset. We chose a case with negligible noise ($\textrm{SNR}=10$), and a difficult case where the noise is of the same magnitude as the signal ($\textrm{SNR}=1$). We generated sample datasets of different sizes, with 100, 1000 and, 10000 samples. 
The high range of these values is representative of typical single-cell flow cytometry experiments. The combination of the different target distribution types, noise distribution types, SNRs, and sample sizes generates a collection of 54 datasets with known ground truth.

\subsection*{Flow cytometry}

E14 mouse embryonic stem cells containing a knock-in fluorescence reporter for the mesodermal transcription factor Brachyury, T/Bra::GFP were used \cite{Fehling:2003us}.
Cells containing the T/Bra::GFP reporter were cultured in ES-Lif (ESL) medium (KnockOut Dulbecco's Modified Eagle Medium (DMEM) supplemented with 10\% fetal bovine serum (FBS), 1x Non-essential aminoacids (NEEA), $50$ U/mL Pen/Strep, 1x GlutaMax, 1x Sodium Pyruvate, $50$ $\mu$M 2-Mercaptoethanol and leukemia inhibitory factor (LIF)). Cells adhered to 0.1\% gelatin-coated (Millipore, ES-006-B) tissue culture-treated 25 cm$^2$ (T25 Corning 353108) plates, and were passaged every second day, as previously described \cite{Turner:2017um}. Cells were kept at 37$^{\circ}$C with 5\% $\mbox{CO}_{2}$, and were routinely tested and confirmed to be free from mycoplasma. Flow cytometry experiments were performed as follows: On day 1, cells were trypsinized and seeded into gelatin-coated 6-well plates (Corning, 353224) to a final density of $\sim 10^5$ cells/well in 3 mL ESL media. On day 2, the media was replaced by first washing twice with 3 mL DPBS$^{+/+}$ (Phosphate buffered saline containing Mg$^{++}$ and Ca$^{++}$, Sigma, D8662) and then adding NDiff227 media (N2B27) (Takara Bio, \#Y40002) \cite{Turner:2017um} with the appropriate combination of Brachyury activators (3 $\mu$M CHIR99, Sigma, SML1046 and/or 25 ng/mL Activin A, Bio-Techne, 338-AC-010).  The same procedure was followed on day 3.  For the control conditions, the corresponding volume of dimethyl sulfoxide (DMSO) was added to the medium. 
Flow cytometry data was acquired on day 2 or day 3, depending on the type of experiment. For the extrinsic noise experiment, cells were incubated with 1 mL of 20 nM CellTracker Green (CMFDA, Thermofisher) for 3 min prior to trypsinization and flow cytometry.
The data was acquired using an LSRIIb flow cytometer, with $2 \times 10^4$ cells being analyzed per condition. DAPI labelling was used to discard death cells and debris. Furthermore, cell doublets were discarded from the analysis. The readout of protein expression (peak of the signal) was obtained through the FITC-A channel.
Measurements were extracted using the FACSDiva software and exported in a Python format for subsequent analysis. 

\subsection*{Software}

The software pipeline presented here, including MCMC for finite and infinite normal mixtures, has been implemented as a Julia package ({\tt scBayesDeconv.jl}).
The source code, with manual compilation and installation instructions, as well as full documentation and a notebook with examples of use, can be obtained publicly from GitHub (\href{https://github.com/dsb-lab/scBayesDeconv.jl}{https://github.com/dsb-lab/scBayesDeconv.jl}).

%
\bibliography{deconv}

\end{document}


\maketitle

\renewcommand{\baselinestretch}{0.2}
\tableofcontents
\renewcommand{\baselinestretch}{1.3}

\section{Biological statement of the problem}

Flow cytometry provides us with measurements of a target signal $T$ in large numbers of single cells.
The target signal is emitted by a fluorophore that reports on the abundance (or activity) of a protein of interest within each cell.
This signal is affected by autofluorescence (which can be considered a source of noise) produced by elements of the cell other than the fluorophore.
Due to the noise, the total signal $C$ measured by the device is not directly $T$, but
\begin{equation}
C = T+\xi
\label{eq:sum}
\end{equation}

\subsection{Deconvolving signal from noise}

We are interested in the case in which we cannot measure both the signal $C$ and the noise $\xi$ independently in the same cell, and thus $T$ in that cell cannot be calculated trivially via Eq.~(\ref{eq:sum}).
This limitation is typical of flow cytometry experiments, in which cells can only be measured once.
In this case, the only information that can be extracted from the device consists of the distributions of the measured signal, $p_c$, and of the background noise by itself, $p_\xi$ (by measuring cells without fluorophore), over large populations of cells (tens of thousands in a typical flow cytometry run).
We can assume the samples to be independent and identically distributed (iid).

If $T$ and $\xi$ and independent of each other, the distributions defined above are related to one another by means of a convolution:
\begin{equation}
p_c(C)=\int_{0}^\infty \!\!\!\int_{0}^\infty \!\!p_T(T)\,p_\xi(\xi)\,\delta(C-(T+\xi))\,d\xi \,dT
= \int_{0}^\infty \!\!p_T(C-\xi)\,p_\xi(\xi)\,d\xi \equiv (p_T*p_\xi)(C)
\label{eq:conv}
\end{equation}

In what follows, we describe a method to extract the distribution $p_T$ of the target variable $T$ from the observed distributions of $C$ and $\xi$ via a deconvolution of Eq.~(\ref{eq:conv}).
The method is applicable to any measurement technique that provides distributions of a signal affected by noise.

\subsection{Generalization to multichannel measurements}

Flow cytometry systems have multiple detectors to measure light at different emission frequencies. This allows to target different proteins in the same cell with different fluorophores, measuring their emission at either the peak frequency (classical flow cytometry) or in a set of frequencies that define the emission spectrum (spectral flow cytometry). Independently of the method, the emission of a fluorophore extends over the spectrum, and hence it spills over the different channels. Considering the additive effect of the emission of different fluorophore markers on the measured channels, equation \eqref{eq:sum} can be more generally stated as
\begin{equation}
c_{k} = \sum_{j=1}^{N_{\rm T}} t_{j}S_{jk}+\xi_{k},\qquad k=1\ldots N_{\rm ch}
\label{eq:sum_generalized}
\end{equation}
where $c_{k}$, $t_{j}$ and $\xi_{k}$ are realizations of the random variables $C$, $T$ and $\xi$ defined in Eq.~(\ref{eq:sum}) above. The subindex $j$ runs over the $N_{\rm T}$ fluorophores (one per target), and the subindex $k$ runs over the $N_{\rm ch}$ channels (so that $\xi_{k}$ represents the contribution of the autofluorescence to channel $k$). The terms $S_{jk}$ define the spillover matrix $\bS$, that quantifies how the fluorophore signals are spread over all the measuring channels.

The spillover matrix can be estimated from single fluorophore controls by using regression methods \cite{roca_autospill_2021}. To that end, one can perform control experiments in which only one fluorophore is present, and measure the resulting signal in all the channels:
\begin{equation}
    c_{ik}^{(j)}=t_{ij}^{(j)}S_{jk}+\xi_{ik}
\end{equation}
where the subindex $j$ now corresponds to the only fluorophore present in the system, the superindex $(j)$ indicates that the experiments were performed in the single-fluorophore condition, and $c_{ik}^{(j)}$ denotes a total signal in channel $k$ coming from cell $i$ when only the fluorophore $j$ is present in the system. This set of equations is underdetermined, as we do not know neither the real target signal $t_{ij}^{(j)}$ nor the components $S_{jk}$ of the spillover matrix. However, if we focus for the moment on the channel that corresponds to our single fluorophore ($k=j$) and impose $S_{jj}=1$, we can write down that $c_{ij}^{(j)}\approx t_{ij}^{(j)}$ (ignoring for now the autofluorescence of that channel). This allows us to establish a set of linear regression problems whose solution enables the estimation of the spillover matrix components for which we have single-fluorophore controls:
\begin{equation}
    c_{ik}^{(j)}\approx c_{ij}^{(j)}S_{jk}+\xi_{ik}
\end{equation}
The spillover coefficients $S_{jk}$ can then be estimated using robust regression techniques that remove the noise coming from outliers \cite{roca_autospill_2021}.

A similar method can be used to reduce the noise coming from the autofluorescence. To that end, one can define an effective “fluorophore signal” $t_{in}$ coming from the autofluorescence, with an associated additional channel $n$ not linked to a real fluorophore used in the sample. If the signal in this channel is correlated with the emission of the autofluorescence $\xi$ over the other channels, we can decompose the autofluorescence signal at every channel $k$ into a regression term and a noise term:
\begin{equation}
    \xi_{ik} \equiv t_{in}S_{nk}+\xi'_{ik} \quad \Longrightarrow \quad c_{ik}=\sum_jt_{ij}S_{jk}+t_{in}S_{nk}+\xi'_{ik}
    \label{eq:channeln}
\end{equation}
The coefficients $S_{nk}$ indicate how the autofluorescence signal is spread over the other channels. The fluorophore signals and the autofluorescence “signal” can be grouped in a single term:
\begin{equation}
    c_{ik}=\sum_{j=1}^{N_{\rm T}+1}t_{ij}\bar{S}_{jk}+\xi'_{ik}
    \label{eq:group}
\end{equation}
where now the sum over $j$ also includes the autofluorescence signal. We can then calculate the spillover coefficients with the method above, by using the additional channel $n$ as an additional "fluorophore control" for the autofluorescence signal. We also note that writing the autofluorescence as a regression problem in Eq.~(\ref{eq:channeln}) will lead in general to a reduction of the noise in all channels:
\begin{equation}
    \text{Var}(\xi'_k)  <  \text{Var}(\xi_k)
\end{equation}
Finally, to obtain the original signal, we can multiply Eq.~(\ref{eq:group}) by the inverse of the spillover matrix:
\begin{equation}
    \bC\bSbar^{-1} = \bT+\bxi'\bSbar^{-1} \equiv \bT+\bxi''
    \label{eq:sum_effective}
\end{equation}
We emphasize that this method requires that the autofluorescence signal at the "noise channel" is sufficiently correlated with its contribution at the other channels, which is not necessarily true \textit{a priori}. 
Also, Eq.~\eqref{eq:sum_effective} shows that the need to deconvolve the noise stands even after the application of the spillover  and autofluorescence corrections proposed in the literature. In fact, the deconvolution method that we propose in this article is compatible with existing methods of autofluorescence correction such as the regression method reviewed above, which can be performed before applying our method to the corrected system \eqref{eq:sum_effective}.

\section{Mathematical statement of the problem}

Since we do not know the exact underlying distributions, we model them as potentially infinite mixtures of normal basis functions. 
The Gaussian mixture models for the target and the noise distributions can be represented as
\begin{subequations}
    \begin{align}
        p(\bt|\bwtar,\set{\bmutar},\set{\bSigmatar})&  = \sum^{K_T}_{i=1} \wtar_i \mathcal{N}\left(\bt|\bmutar_i, \bSigmatar_i\right)\label{eq:pT}\\
        p(\bxi|\bwn,\set{\bmun},\set{\bSigman})& = \sum^{K_\xi}_{i=1} \bwn_i \mathcal{N}\left(\bxi|\bwn_i, \bSigman_i\right),\label{eq:pxi}
    \end{align}
    \label{eq:mixtures}
\end{subequations}
where $\mathcal{N}(\bx|\bmu_i, \bSigma_i)$ denotes a normal distribution on the variable vector $\bx$ (whose components are the different measurement channels), with mean $\bmu_i$ and variance $\bSigma_i$.
$K_T$ and $K_\xi$ represent the number of bases used to describe each distribution, $\bm{\omega}=\{\omega_i:i=1\ldots K\}$ are their weights, and $\set{\bm{\mu}}=\{\bmu_i:i=1\ldots K\}$ and $\set{\bSigma}=\{\bSigma_i:i=1\ldots K\}$ are their characteristic parameters.
Note that we have removed the subindices from the distributions of the target and the noise, since according to the mixture representation, $\target{p}$ and $\noise{p}$ depend exclusively on the parameters defined above.
 
The mixture decomposition defined in Eqs.~(\ref{eq:mixtures}) allows for a flexible and robust representation of unknown and generic distributions. Moreover, exploiting the fact that the convolution of two Gaussian distributions is Gaussian, we have an analytical expression for the distribution of the total variable $\bC$:
\begin{equation}
    p(\bc|\bphitar,\bphin)= \sum_{i=1}^{K_T}\sum_{j=1}^{K_\xi} \wtar_j\wn_j \N\left(\bc\left|\bmutar_i+\bmun_j, \bSigmatar_i+\bSigman_j\right.\right)
\label{eq:pc}
\end{equation}
where $\bphitar=\{\bwtar,\bmutar,\bSigmatar\}$ and $\bphin=\{\bwn,\bmun,\bSigman\}$ represent all the parameters of the target and noise distributions, from which we can parametrize the distribution of the total measured signal.
We note that the combination of normal basis function $\N$ in Eq.~(\ref{eq:pc}) above is invariant under changes in the individual basis functions of $T$ and $\xi$, provided the sums of means, $\bmutar+\bmun,$ and variances, $\bSigmatar+\bSigman$, are constant.

As discussed in the main text, according to Bayes' rule, the posterior distribution that represents the probability of the parameters given the data is
\begin{equation}
p(\bphitar,\bphin|\set{\mathbf{c}},\set{\boldsymbol\xi})\propto \Big[p(\set{\mathbf{c}}|\bphitar,\bphin)p(\set{\bxi}|\bphin)\Big]
\Big[p(\bphin)p(\bphitar)\Big],
\label{eq:b_rule}
\end{equation}
where $\set{\mathbf{c}}=\{\bc_i:i=1\ldots N_c\}$ and $\set{\bm\xi}=\{\bxi_i:i=1\ldots N_\xi\}$ represent the samples of the total signal and noise, respectively, with the different components of $\bc_i$ and $\bxi_i$ correspond to the different channels of the cytometer.
In Eq.~(\ref{eq:b_rule}), the first bracket on the right-hand side corresponds to the likelihood, namely the joint probability of observing the data given the parameters, which we can redefine as
\begin{equation}
\mathcal{L}=p(\set{\mathbf{c}}|\bphitar,\bphin)p(\set{\bm\xi}|\bphin)=\prod_i^{N_c} p(\bc_i|\bphitar,\bphin)\prod_j^{N_\xi} p(\bxi_j|\bphin),
\label{eq:likelihood}
\end{equation}
The form of the likelihood (\ref{eq:likelihood}) breaks the symmetry between the target and noise signals, and thus lifts the above-mentioned degeneracy between their parameters exhibited by Eq.~\eqref{eq:pc}. 
The second bracket in the right-hand side of Eq.~(\ref{eq:b_rule}), in turn, corresponds to the prior distributions of all the parameters of the problem, which we define in what follows.

\subsection{Approximated decomposition of the posterior into two separate problems}
\label{Posterior_approximation}

It is worth noting that the posterior distribution \eqref{eq:b_rule} can be decomposed as follows:
\begin{equation}
    p(\bphitar,\bphin|\set{\bc},\set{\boldsymbol\xi}) = p(\bphitar|\bphin,\set{\bc})p(\bphin|\set{\mathbf{c}},\set{\boldsymbol\xi})
\end{equation}
where we have removed the dependency of the noise signal in the first term of the right-hand side, since only the total signal $\bc$ defines the parameters of the target distribution, as can be seen from the likelihood \eqref{eq:likelihood}. The second term, on the other hand, is conditioned by both the noise and the total signal. Since we usually have as much data for the noise signal $\set{\boldsymbol\xi}$ as for the signal of interest $\set{\bc}$, we can approximately consider that the posterior distribution of the noise mixture parameter $\bphin$ is well represented by the noise data alone: 
\begin{equation}
    p(\bphitar,\bphin|\set{\mathbf{c}},\set{\boldsymbol\xi}) \approx p(\bphitar|\bphin,\set{\mathbf{c}})p(\bphin|\set{\bxi})
    \label{eq:b_rule_approx}
\end{equation}
With this approximation, the problem can be decomposed in two separate subproblems: first, finding the probability distribution of the parameters $\bphin$ of the noise mixture; and second, finding the distribution of the parameters $\bphitar$ of the convolution mixture, conditioned on the noise mixture parameters.

In the following section, we go over the mathematical details of the probability distributions that will allow us to sample from the posterior distribution.

\section{Relevant probability distributions}

In this section, we derive the main expressions that we need to sample our model. Our aim is to have a self-contained derivation of the sampling process of the posterior of a Gaussian normal mixture.

\subsection{Multivariate normal distribution with unknown mean and error}
\label{multivariate_normal_distribution}

Here we derive the main results of interest on multivariate normal distributions, which we will use when sampling the posterior distribution using normal bases.

\subsubsection{Likelihood}

The multivariate normal distribution for a set of independent identical samples (iid) $\set{\bx}$ has the form, up to a scaling parameter,
\begin{equation}
    \mathcal{N}(\set{\bx};\bmu,\bSigma) \propto \Det{\bSigmaInv}^{n/2}\Normal{\sum_{i=1}^n(\bx_i-\bmu)^T\bSigmaInv(\bx_i-\bmu)}
\end{equation}
where $\mu$ is the mean and $\bSigmaInv$ is the precision parameter and $n$ is the number of cells being measured. The precision parameter relates to the covariance matrix as
\begin{equation}
 \bSigmaInv = \bSigma^{-1}
\end{equation}

It is convenient to rearrange the distribution to make explicit the dependency of the summary statistics:
\begin{multline}
    \mathcal{N}(\set{\bx};\bmu,\bSigma) \propto \Det{\bSigmaInv}^{n/2}\Normal{\sum_i(\bx_i-\bmu)^T\bSigmaInv(\bx_i-\bmu)}\\
    \propto \Det{\bSigmaInv}^{n/2}\Normal{\sum_i(\bx_i-\mbx)^T\bSigmaInv(\bx_i-\mbx)+n(\mbx-\bmu)^T\bSigmaInv(\mbx-\bmu)} \\   
    \propto \Det{\bSigmaInv}^{n/2}\Normal{n\Tr{\bs{x}\bSigmaInv}+n(\mbx-\bmu)^T\bSigmaInv(\mbx-\bmu)}    
\label{eq:mn_statistics}
\end{multline}
where the summary statistics are the mean,
\begin{equation}
    \mbx=\frac{1}{n}\sum_i\bx_i
\end{equation}
and the covariance,
\begin{equation}
    \bs{x}=\frac{1}{n}\sum_i(\bx_i-\mbx)(\bx_i-\mbx)^T
\end{equation}

\subsubsection{Conjugate prior}

A convenient prior for the multivariate distribution is a conjugate prior that allows us to obtain the analytic form of the other distributions:
\begin{align}
p(\bmu|\bSigma) &=  \N(\bmu;\bmu_0,\kappa_0\bSigma)\\
p(\bSigma) &=  \InvW(\bSigma;\bSigma_0,\nu_0)   
\label{eq:mn_conjugate_prior}
\end{align}
where $\InvW$ is the inverse Wishart distribution. The hyperparameter $\kappa_0$ represents our confidence in the estimation of the mean: the higher $\kappa_0$ is, the closer to the mean we will be.

The inverse Wishart distribution has the following shape, up to scaling terms:
\begin{equation}
    \InvW(\bm{X};\bm{\Sigma},\nu) \propto \frac{\Det{\bm{\Sigma}}^{(\nu)/2}}{\Det{\bm{X}}^{(\nu+p+1)/2}}e^{-\Tr{\bm{\Sigma}\bm{X}^{-1}}/2}
\end{equation}
where $p$ is the number of dimensions. The parameter $\bm{\Sigma}$ is the correlation matrix and $\nu$ represents our confidence in the estimation of the $\bm{\Sigma}$ correlation matrix. 

\subsubsection{Posterior distribution}

The posterior distribution will have the following shape:
\begin{multline}
    p(\bmu,\bSigma|\set{\bx}) \propto p(\set{\bx}|\bmu,\bSigma)p(\bmu|\bSigma,\kappa_0,\bmu_0)p(\bSigma|\bSigma_0,\nu_0) \\ 
    \propto \Det{\bSigmaInv}^{n/2}\Normal{n\Tr{\bs{x}\bSigmaInv}+n(\mbx-\bmu)^T \bSigmaInv (\mbx-\bmu)}\times \\
    \Det{\bSigmaInv}^{1/2}\Normal{\kappa_0(\bmu-\bmu_0)^T \bSigmaInv (\bmu-\bmu_0)}\Det{\bSigmaInv}^{(\nu_0+p+1)/2}\Normal{\Tr{\bSigma_0\bSigmaInv}}
\end{multline}
We can use the posterior distribution to obtain different distributions of relevance. 

\subsubsection{Conditional distribution: mean}
\label{conditional_distribution_mean}

Retaining the terms involving the mean $\bmu$, the conditional distribution of the mean takes the form of a multivariate distribution:
\begin{align}
    p(\bmu|\bSigma,\set{\bx}) &\propto \Normal{n(\mbx-\bmu)^T \bSigmaInv (\mbx-\bmu)+\kappa_0(\bmu-\bmu_0)^T \bSigmaInv (\bmu-\bmu_0)} \nonumber\\
    &\propto \Normal{-2(n\mbxT+\kappa_0\bmu_0)\bSigmaInv\bmu+(n+\kappa_0)\bmuT \bSigmaInv\bmu} \nonumber\\
    &\propto \Normal{(\bmu-\bmueff)^T\bSigmaInveff(\bmu-\bmueff)}
    \quad \propto \N(\bmu|\bmueff,\bSigmaInveff)
    \label{eq:cd_mean}
\end{align}
where in the last step we completed the squares. This multivariate normal distribution has effective parameters
\begin{equation}
\bmueff = \frac{n\mbx+\kappa_0\bmu_0}{n+\kappa_0}, \qquad \bSigmaInveff = (n+\kappa_0)\bSigmaInv
\end{equation}
where  $\bmueff$ is a weighted version between the mean statistic and the prior mean, and $\bSigmaInveff$ is an effective precision parameter. The $\kappa_0$ factor indicates how close we are from the prior mean. 

\subsubsection{Conditional distribution: covariance}
\label{conditional_distribution_covariance}

Retaining the terms involving the covariance matrix (and its inverse, the precision matrix), the conditional distribution can be shown to take the form of an inverse Wishart distribution:
\begin{align}
    &p(\bSigma|\bmu,\set{\bx}) \propto \nonumber \\
    &\Det{\bSigmaInv}^{(n+\nu_0+p+2)/2}
    \Normal{n\Tr{\bs{x}\bSigmaInv}+n(\mbx-\bmu)^T \bSigmaInv (\mbx-\bmu)+\kappa_0(\bmu-\bmu_0)^T \bSigmaInv (\bmu-\bmu_0)+\Tr{\bSigma_0\bSigmaInv}} \nonumber\\
    &\propto \Det{\bSigmaInv}^{(n+\nu_0+p+2)/2}
    \Normal{\Tr{\left(n\bs{x}\bSigmaInv+n(\mbx-\bmu)(\mbx-\bmu)^T+\kappa_0(\bmu-\bmu_0)(\bmu-\bmu_0)^T+ \bSigma_0\right)\bSigmaInv}} \nonumber\\
    &\propto \Det{\bSigmaInv}^{(\neff+p+1)/2}
    \Normal{\Tr{\bSigmaeff\bSigmaInv}} \quad\propto \InvW(\bSigma|\bSigmaeff,\neff)
    \label{eq:cd_covariance}
\end{align}
where the effective parameters are
\begin{equation}
    \neff = n+\nu_0+1
\end{equation}
\begin{equation}
    \bSigmaeff = n\bs{x}+n(\mbx-\bmu)(\mbx-\bmu)^T+\kappa_0(\bmu-\bmu_0)(\bmu-\bmu_0)^T+ \bSigma_0
\end{equation}
In this last expression, it is worth noting that each term represents a different kind of uncertainty: the first term is the uncertainty coming from the mean statistic, the second corresponds to the uncertainty with which the mean statistic represents the actual mean, the third term is the uncertainty coming from the prior mean, and the last term is the uncertainty coming from the prior itself.

\subsubsection{Marginal distribution: covariance}
\label{marginal_distribution_variance}

One additional distribution that we will need is the marginal distribution of the variance:
\begin{align}
    &p(\bSigma|\set{\bx}) = \int p(\bSigma,\bmu|\set{\bx_i}_i)d\bmu
    \propto \int \Det{\bSigmaInv}^{n/2}\Normal{n\Tr{\bs{x}\bSigmaInv}+n(\mbx-\bmu)^T \bSigmaInv (\mbx-\bmu)} \nonumber \\
    &\quad\times\Det{\bSigmaInv}^{1/2}\Normal{\kappa_0(\bmu-\bmu_0)^T \bSigmaInv (\bmu-\bmu_0)}\Det{\bSigmaInv}^{(\nu_0+p+1)/2}\Normal{\Tr{\bSigma_0\bSigmaInv}}d\bmu \nonumber\\
    &\quad\propto \Det{\bSigmaInv}^{(n+\nu_0+p+2)/2}\Normal{n\Tr{\bs{x}\bSigmaInv}+\Tr{\bSigma_0\bSigmaInv}} \nonumber\\
    &\quad\times\left[\int \Normal{\underbrace{\kappa_0(\bmu-\bmu_0)^T \bSigmaInv (\bmu-\bmu_0)+n(\mbx-\bmu)^T \bSigmaInv (\mbx-\bmu)}_{(*)}d\bmu}\right]
\end{align}

Reorganizing the elements in $(*)$,
\begin{align}
    (*) = (\kappa_0+n)\bmuT\bSigmaInv\bmu - 2\bmuT\bSigmaInv(\kappa_0\bmu_0+n\mbx) + (n\mbxT\bSigmaInv\mbx+\kappa_0\bmu_0\bSigmaInv\bmu_0) \\
    = (\bmu-\bmueff)^T\bSigmaInveff(\bmu-\bmueff)\underbrace{-\bmueffT\bSigmaInveff\bmueff+ (n\mbxT\bSigmaInv\mbx+\kappa_0\bmu_0\bSigmaInv\bmu_0),}_{(**)}
\end{align}
where we have defined,
\begin{align}
    \bmueff &= \frac{\kappa_0\bmu_0+n\mbx}{\kappa_0+n}\\
    \bSigmaInveff &= (\kappa_0+n)\bSigmaInv,
\end{align}
we can further regroup the part in $(**)$.
\begin{multline}
    (**) = \frac{-\kappa_0^2\bmuT_0\bSigmaInv\bmu_0-n^2\mbxT\bSigmaInv\mbx-2\kappa_0n\bmuT_0\bSigmaInv\mbx+n^2\mbxT\bSigmaInv\mbx+\kappa_0^2\bmu_0\bSigmaInv\bmu_0+n\kappa_0\mbxT\bSigmaInv\mbx+\kappa_0n\bmu_0\bSigmaInv\bmu_0}{\kappa_0+n}\\
    \frac{\kappa_0n(\mbxT\bSigmaInv\mbx-2\bmuT_0\bSigmaInv\mbx+\bmu_0\bSigmaInv\bmu_0)}{\kappa_0+n}=\frac{\kappa_0n}{\kappa_0+n}(\mbx-\bmu_0)^T\bSigmaInv(\mbx-\bmu_0)
\end{multline}
    
Inserting these terms
\begin{align}
    p(\bSigma|\set{\bx}) &\propto 
    \Det{\bSigmaInv}^{(n+\nu_0+p+2)/2}\Normal{\Tr{\left(n\bs{x}+\bSigma_0+\frac{\kappa_0n}{\kappa_0+n}(\mbx-\bmu_0)(\mbx-\bmu_0)^T\right)\bSigmaInv}} \nonumber\\
    &\times\left[\int \Normal{(\bmu-\bmueff)^T\bSigmaInveff(\bmu-\bmueff)}d\bmu\right]
\end{align}
now we can calculate the integral as a multivariate normal:
\begin{align}
    &p(\bSigma|\set{\bx})\propto  \nonumber\\
    &\propto\Det{\bSigmaInv}^{(n+\nu_0+p+2)/2}\Normal{\Tr{\left(n\bs{x}+\bSigma_0+\frac{\kappa_0n}{\kappa_0+n}(\mbx-\bmu_0)(\mbx-\bmu_0)^T\right)\bSigmaInv}}
    \Det{\bSigmaInveff}^{-1/2} \nonumber\\
    &\propto \Det{\bSigmaInv}^{(n+\nu_0+p+1)/2}\Normal{\Tr{\bSigmaeff\bSigmaInv}} \propto \InvW(\bSigma|\bSigmaeff,\neff)
\end{align}
where in the last step we use the fact that $\det{\bSigmaInveff}\propto\det{\bSigmaInv}$ and the effective parameters of the inverse Wishart distribution are
\begin{align}
\neff &= n+\nu_0+1\\
\bSigmaeff &= n\bs{x}+\frac{\kappa_0n}{\kappa_0+n}(\mbx-\bmu_0)(\mbx-\bmu_0)^T +\bSigma_0
\end{align}

\subsubsection{Posterior predictive distribution}
\label{posterior_predictive_distribution}

One last distribution of interest is the probability of a new data point given the already observed set of observations:
\begin{align}
    p(\by|\set{\bx}) &= \iint p(\by|\bmu,\bSigma)p(\bmu,\bSigma|\set{\bx_i}_i)d\bmu d\bSigma
    \quad\propto \iint
    \Det{\bSigmaInv}^{(n+\nu_0+p+3)/2} \nonumber\\
    &\times\Normal{\underbrace{(\by-\bmu)^T \bSigmaInv (\by-\bmu)+n(\mbx-\bmu)^T \bSigmaInv (\mbx-\bmu)+\kappa_0(\bmu-\bmu_0)^T \bSigmaInv (\bmu-\bmu_0)}_{(*)}}\nonumber\\
    &\times\Normal{\Tr{(n\bs{x}+ \bSigma_0)\bSigmaInv}}d\bmu d\bSigma
\end{align}

We can rearrange the elements in $(*)$, completing squares as we did when calculating the mean conditional distribution (section\ref{conditional_distribution_mean}):
\begin{equation}
    (*) = \underbrace{n\mbxT\bSigmaInv\mbx + \byT \bSigmaInv \by + \kappa_0\bmuT_0\bSigmaInv\bmu_0  - \bmueff\bSigmaInveff\bmueff^\mu}_{(**)} +(\bmu-\bmueff^\mu)^T\bSigmaInveff(\bmu-\bmueff^\mu),
\end{equation}
where we have retained all the terms involving $\bSigmaInv$ and $\bmu$ this time, as they are necessary for integrating out. The effective parameters are now
\begin{equation}
    \bmueff^\mu = \frac{n\mbx+\kappa_0\bmu_0+\by}{n+\kappa_0+1}
\end{equation}
\begin{equation}
    \bSigmaInveff^\mu = (n+\kappa_0+1)\bSigmaInv
\end{equation}
We can further rearrange the terms in $(**)$ to obtain an expression with the $y$ in quadrature:
\begin{align}
    &(**) = n\mbxT\bSigmaInv\mbx + \byT \bSigmaInv \by + \kappa_0\bmuT_0\bSigmaInv\bmu_0  - \frac{1}{n+\kappa_0+1}(n\mbx+\kappa_0\bmu_0+\by)^T\bSigmaInv(n\mbx+\kappa_0\bmu_0+\by)\nonumber\\
    &= \frac{1}{n+\kappa_0+1}\left[(n\mbxT\bSigmaInv\mbx + \byT \bSigmaInv \by + \kappa_0\bmuT_0\bSigmaInv\bmu_0)(n+\kappa_0+1)  - (n\mbx+\kappa_0\bmu_0+\by)^T\bSigmaInv(n\mbx+\kappa_0\bmu_0+\by)\right]\nonumber\\
    &= \frac{1}{n+\kappa_0+1}\left[n(\by-\mbx)^T\bSigmaInv(\by-\mbx)+\kappa_0(\by-\bmu_0)^T\bSigmaInv(\by-\bmu_0)+n\kappa_0(\bmu_0-\mbx)^T\bSigmaInv(\bmu_0-\mbx)\right]\nonumber\\
    &= \frac{(n+\kappa_0)}{n+\kappa_0+1}(\by-\bmueff^y)^T\bSigmaInv(\by-\bmueff^y)+\frac{(\frac{n\kappa_0}{n+\kappa_0}+n\kappa_0)}{n+\kappa_0+1}(\bmu_0-\mbx)^T\bSigmaInv(\bmu_0-\mbx)\nonumber\\
    &= m^0(\by-\bmueff^y)^T\bSigmaInv(\by-\bmueff^y)+m^1(\bmu_0-\mbx)^T\bSigmaInv(\bmu_0-\mbx)
\end{align}
where in the second to the third step we group by quadrature and complete squares to group the terms with $y$, and then group by quadrature the terms on the left. The effective parameters are
\begin{equation}
    \bmueff^y = \frac{(n\mbx+\kappa_0\bmu_0)}{n+\kappa_0},\qquad
    m^0 = \frac{n+\kappa_0}{n+\kappa_0+1},\qquad
    m^1 = \frac{\frac{n\kappa_0}{n+\kappa_0}+n\kappa_0}{n+\kappa_0+1}
\end{equation}

We can now insert the obtained term in the original expression:
\begin{multline}
    p(\by|\set{\bx})\propto \iint
    \Det{\bSigmaInv}^{(n+\nu_0+p+3)/2}\times \\
    \Normal{m^0(\by-\bmueff^y)^T\bSigmaInv(\by-\bmueff^y)+m^1(\bmu_0-\mbx)^T\bSigmaInv(\bmu_0-\mbx) +\Tr{n\bs{x}\bSigmaInv} +\Tr{\bSigma_0\bSigmaInv}}\times \\
    \Normal{(\bmu-\bmueff^\mu)^T\bSigmaInveff^\mu(\bmu-\bmueff^\mu)} d\bmu d\bSigma\\
    = \int
    \Det{\bSigmaInv}^{(n+\nu_0+p+3)/2}\Normal{\Tr{\bSigmaeff\bSigmaInv}}\times\\
    \left[\int \Normal{(\bmu-\bmueff^\mu)^T\bSigmaInveff^\mu(\bmu-\bmueff^\mu)} d\bmu\right] d\bSigma
    \label{eq:ppd}
\end{multline}
where the effective covariance has the form,
\begin{equation}
    \bSigmaeff = m^0(\by-\bmueff^y)(\by-\bmueff^y)^T+m^1(\bmu_0-\mbx)(\bmu_0-\mbx)^T +n\bs{x} + \bSigma_0
\end{equation}

The mean parameter in \eqref{eq:ppd} only appears in the term in brackets, so we can integrate it in a straightforward manner as a multivariate normal integral:
\begin{multline}
    p(\by|\set{\bx})\propto \int
    \Det{\bSigmaInv}^{(n+\nu_0+p+3)/2}\Normal{\Tr{\bSigmaeff\bSigmaInv}}\Det{\bSigmaInveff}^{-1/2} d\bSigma\\
    \propto \int \Det{\bSigmaInv}^{(n+\nu_0+p+2)/2}\Normal{\Tr{\bSigmaeff\bSigmaInv}}d\bSigma
\end{multline}
where $\Det{\bSigmaInveff^\mu}\propto\Det{\bSigmaInv}$. The last integral is an inverse Wishart that we can integrate directly, leading to
\begin{equation}
    p(\by|\set{\bx})\propto \Det{\bSigmaeff}^{-(n+\nu_0+1)/2}
    \label{eq:multivariate_t_distribution_1}
\end{equation}

We can now reorganize the effective covariance. If we define the matrix
\begin{equation}
    \bSigmaeff^y = \frac{1}{m^0}\left(m^1(\bmu_0-\mbx)(\bmu_0-\mbx)^T +n\bs{x} + \bSigma_0\right),
\end{equation}
we can rewrite the effective covariance $\bSigmaeff$ as
\begin{equation}
    \bSigmaeff=\left((\by-\bmueff^y)(\by-\bmueff^y)^T\bSigmaInveff^y+\mathbb{I}\right)m_0\bSigmaeff^y,
\end{equation}
Insert this expression in equation \eqref{eq:multivariate_t_distribution_1} we obtain
\begin{align}
    p(\by|\set{\bx})&\propto\Det{\left((\by-\bmueff^y)(\by-\bmueff^y)^T\bSigmaInveff^y+\mathbb{I}\right)n_0\bSigmaeff^y}\nonumber\\
    &\propto \Det{(\by-\bmueff^y)(\by-\bmueff^y)^T\bSigmaInveff^y+\mathbb{I}}^{-(n+\nu_0+1)/2}
    =|1+(\by-\bmueff^y)^T\bSigmaInveff^y(\by-\bmueff^y)|^{-(n+\nu_0+1)/2}\nonumber\\
    &=|1+\frac{1}{\nueff}(\by-\bmueff^y)^T(\nueff\bSigmaInveff^y)(\by-\bmueff^y)|^{-(\nueff+p)/2}=\TStudent{\nueff}(\by;\bmueff^y,\bSigmaeff^y/\nueff)
\end{align}
which is a multivariate T-distribution with
\begin{align}
    m^0 &= \frac{n+\kappa_0}{n+\kappa_0+1},\quad m^1 = \frac{\frac{n\kappa_0}{n+\kappa_0}+n\kappa_0}{n+\kappa_0+1},
    \quad \nueff = n+\nu_0+1-p\\
    \bmueff^y &= \frac{(n\mbx+\kappa_0\bmu_0)}{n+\kappa_0}\\
    \bSigmaeff^y &= \frac{1}{m^0}\left(m^1(\bmu_0-\mbx)(\bmu_0-\mbx)^T +n\bs{x} + \bSigma_0\right)
\end{align}

\subsection{Finite mixture distributions}

We derive in this section the statistics relevant to mixture models. 

\subsubsection{Likelihood}
\label{fmm_likelihood}

The likelihood of a finite mixture model with $K$ components has the form
\begin{equation}
    p(\set{\bx}|\set{\bphi},\bw)=\prod_i\left(\sum_j^K \w_j p(\bx_i|\bphi_j)\right),
    \label{eq:fmm_sum}
\end{equation}
where the vector sets $\set{\bx}=\{\bx_i:i=1\ldots N\}$ and $\set{\bphi}=\{\bphi_j:j=1\ldots K\}$ run over the number of samples (cells) $N$ and the number of mixture components $K$, respectively (in what follows we use the subindices $i$ and $j$ with those two distinct meanings)\footnote{We remind the reader that the dimension of the vectors $\bx_i$ and $\bphi_j$ is equal to the number of measurement channels.}. The set of parameters $\bw$ are called the weights of the mixture model, and $p(\set{\bx}|\bphi_j)$ are the base distributions. We can extend this model to introduce a set of hidden indicator variables $\set{\bz}$, defined as
\begin{equation}
    z_{ij} = \delta_{jk_i}
\end{equation}
for some $k_i \in \set{1,...,K}$, and where $\delta_{ij}$ is the Kronecker delta. This variable basically tells from which distribution $p(\set{\bx}|\bphi_i)$ the variable came from. Using this set of hidden variables, our model \eqref{eq:fmm_sum} can be rewritten as
\begin{equation}
    p(\set{\bx},\set{\bz}|\set{\bphi},\bw)=\prod_i^N\prod_j^K p(\bx_i|\bphi_j)^{z_{ij}}\w_j^{z_{ij}}
    \label{eq:fmm}
\end{equation}

It is straightforward to see that, if we take the marginal distribution over the hidden variables, we recover the original distribution:
\begin{equation}
p(\set{\bx}|\set{\bphi},\bw) =  \prod_{i}^N\left(\sum_{j}^K p(\bx_i,z_{ij}=1|\bphi_j,\bw)\right) =  \prod_{i}^N\left(\sum_{j}^Kp(\bx_i|\bphi_j)\w_{j}\right)
\end{equation}
where only when the indicator variable is one of the corresponding term survives. 

The use of indicator variables makes it possible to compute analytically the posterior distribution of the mixture model, as the base distributions are now in product form. The hidden indicator variables are not known, thus we will have to sample from them as well.

\subsubsection{Conjugate prior}

A conjugate prior for the mixture model is
\begin{equation}
    p(\bw|\balpha) = \Dirichlet(\bw|\balpha)
\end{equation}
where the Dirichlet distribution has the form
\begin{equation}
    \Dirichlet(\bw|\balpha) \propto \prod_i^K \w_i^{\alpha_i-1}
    \label{eq:fmm_prior}
\end{equation}

The hyperprior parameters $\alpha$ are usually set to be symmetrical and to scale with the number of mixture components:
\begin{equation}
    \alpha_i = \alpha/K\quad\forall i
\end{equation}
In this way, the prior distribution only depends on one hyperparameter $\alpha$ that indicates the strength from the uniform weights.

\subsubsection{Posterior distribution}

Putting together the likelihood and the prior distribution, the posterior of the mixture model is
\begin{equation}
    p(\set{\bphi},\bw|\set{\bx},\set{\bz})\propto \prod_j^K\left(\prod_i p(\bx_i|\bphi_j)^{z_{ij}}\w_j^{z_{ij}}\right)\w_j^{\alpha/K-1}p(\bphi_j^0),
    \label{eq:fmm_posterior}
\end{equation}
where the last term is the set of priors for each base distribution.

\subsubsection{Conditional distribution: weights}
\label{conditional_distribution_weights}

Taking the terms from the posterior that involve the weights:
\begin{equation}
    p(\bw|\set{\bz}) \propto \prod_j^K\left(\prod_i \w_j^{z_{ij}}\right)\w_j^{\alpha/K-1} = \Dirichlet(\bw|\bneff)
    \label{eq:cd_weights}
\end{equation}
where $\neff_j = \sum_i z_{ij}+\alpha$.

\subsubsection{Conditional distribution: indicator variables}
\label{conditional_distribution_indicator_variables}

As we already mentioned, the indicator variables are not known, so we have to sample from them too. As we are considering identically independent samples, we can obtain the conditional distribution from each indicator variable independently as
\begin{equation}
    p(\bc_i|\bw,\set{\bphi},\bx_i) \propto \prod_j^K p(\bx_i|\bphi_j)^{z_{ij}}\w_j^{z_{ij}} = \Multinomial(\bz_i;1,\set{p(\bx_i|\bphi_j)\w_j}_j)
    \label{eq:cd_indicator}
\end{equation}
It is worth noting that the indicator will be sampled from a particular base distribution for the weights of that base, but also by how well that sample lies inside the base distribution.

\subsubsection{Conditional distribution: base distribution parameters}

Finally, because the base distributions are in product form due to the introduction of the indicator variables, the parameters of each base can be computed independently as
\begin{equation}
    p(\bphi_j|\set{\bx},\set{\bz}) \propto p(\set{\bx}_{j}|\bphi_j)p(\bphi^0_j)
\end{equation}
where $\set{\bx}_{j}$ represents the subsample of cells whose indicator variable belongs to the corresponding mixture component.

\subsection{Infinite mixture distributions}

In this section we define basic results from infinite Dirichlet processes (for an insightful tutorial see \cite{li_tutorial_2019}) that allow us to consider infinite mixtures. 

\subsubsection{Taking the infinite limit}

In order to take the limit to infinite clusters, we need to remove the dependence from the number of clusters in a mixture model. For that, we need to remove the dependence on the weights of our probability distribution. Consider for the moment a basis distribution that is uniform in space. The joint probability distribution conditioned on the priors would be
\begin{equation}
    p(\set{\bz},\bw|\alpha)=p(\set{\bz}|\bw)p(\bw|\alpha)
\end{equation}
where the first term in the right-hand side is the likelihood of the indicator variables as in \eqref{eq:fmm}, also given in \eqref{eq:cd_indicator}, which is a multinomial distribution. The second term is the prior distribution of the mixture distribution \eqref{eq:fmm_prior}, which is a Dirichlet distribution. From this expression, we can calculate the marginal distribution of the indicator variables conditioned to the prior parameter:
\begin{align}
    p(\set{\bz}|\balpha)&=\int p(\set{\bz}|\bw)p(\bw|\balpha)d\bw 
    = \int \prod_i\left(\prod_j\w_j^{z_{ij}}\right)\frac{\Gamma(\alpha)}{\Gamma(\alpha/K)^K}\prod_j\w_j^{\alpha/K-1}d\bw
    \nonumber \\
    &= \frac{\Gamma(\alpha)}{\Gamma(\alpha/K)^K}\int \prod_j\w_j^{n_j+\alpha/K-1}d\bw
    = \frac{\Gamma(\alpha)}{\Gamma(\alpha/K)^K}\frac{\prod_j\Gamma(n_j+\alpha/K)}{\Gamma(n+\alpha/K)}
    \label{eq:imm_derivation_1}
\end{align}
where the third equality makes use of the statistic $n_j=\sum_iz_{ij}$. The integral can be identified as a multinomial distribution without the scaling factor.

The expression above still contains explicitly the dependence on the number of components in the mixture $K$, and thus its limit $K\to\infty$ cannot be computed in a straightforward manner. To have a more amenable expression, let us consider that we take out the sample $l$ and reassign it to a new cluster. The probability of this sample to be assigned to any of the clusters, conditioned on all the other indicator variables, is
\begin{align}
    p(z_{lk}=1|\set{\bz}_{\neg l},\alpha)
    &= \frac{p(\set{\bz}_{\neg l},z_{lk}=1|\alpha)}{p(p(\set{\bz}_{\neg l}|\alpha)}
    = \frac{\frac{\Gamma(\alpha)}{\Gamma(\alpha/K)^K}\frac{\prod_j\Gamma(n_{j|l}+\delta_{jk}+\alpha/K)}{\Gamma(n+\alpha/K)}}{\frac{\Gamma(\alpha)}{\Gamma(\alpha/K)^K}\frac{\prod_j\Gamma(n_{j|l}+\alpha/K)}{\Gamma(n+\alpha/K-1)}}\nonumber\\
    &= \frac{\Gamma(n+\alpha/K-1)}{\Gamma(n+\alpha/K)}\prod_j\frac{\Gamma(n_{j|l}+\delta_{jk}+\alpha/K)}{\Gamma(n_{j|l}+\alpha/K)}
    =\frac{n_{k|l}+\alpha/K}{n+\alpha-1},
    \label{eq:imm_derivation_2}
\end{align}
where have made explicit the range of cells over which the samples $\set{\bz}$ are taken in each case. To that end, we define
$\set{\bz}_{\neg l}\equiv\set{\bz}_{i\in(1,...,l-1,l+1,...n)}$
and $n_{\neg l}\equiv\sum_{i\in(1,...,l-1,l+1,...n)}z_{ij}$ to refer to the subsets that contain all elements except $l$. With this expression, it is straightforward to take the limit to infinite components:
\begin{equation}
    p_\infty(z_{lk}=1|\set{\bz}_{\neg l},\alpha) = \lim_{K\rightarrow \infty} p(z_{lk}=1|\set{\bz}_{\neg l},\alpha) = \frac{n_{\neg l}}{n+\alpha-1}
\end{equation}

We now have the probability that a sample is assigned to a base that has other indicator variables. Let us now consider for the moment that we have $K$ bases with assigned samples. The probability that a sample is assigned to a new base will be:
\begin{align}
    p_\infty(\bz_{l,K+1}=1|\set{\bz}_{\neg l},\alpha)
    &= 1 - \sum_{j=1}^Kp_\infty(z_{lk}=1|\set{\bz}_{\neg l},\alpha)\nonumber\\
    &= 1 - \sum_{j=1}^K\frac{n_{k|l}}{n+\alpha-1}
    = 1 - \frac{n-1}{n+\alpha-1}= \frac{\alpha}{n+\alpha-1} 
\end{align}
There is a non-zero probability that the cell will be assigned to a new base that was not populated before, and the probability of populating this new cluster will depend on the hyperparameter $\alpha$.

\subsubsection{Adding a non-uniform basis}
\label{section_imm_complete}

The results above have been derived considering a uniform basis distribution. In the most general case, the basis will be non-uniform. From Eq.~\eqref{eq:fmm} it is very easy to see that if we had a non-uniform distribution, the corresponding term will drop out of the integral \eqref{eq:imm_derivation_1}. Proceeding in the same way as before, all the terms in \eqref{eq:imm_derivation_2} will cancel out, except $p(x_l|\bphi_k)$. 

The probabilities of a new assignation will be
\begin{equation}
    p_\infty(z_{lk}=1|\set{\bz}_{\neg l},\set{\bphi},\alpha) \propto \frac{n_{\neg l}}{n+\alpha-1}p(\bx_i|\bphi_k)
\end{equation}
for a basis with a populated sample and
\begin{equation}
    p_\infty(z_{l,K+1}=1|\set{\bz}_{\neg l},\set{\bphi},\alpha) \propto \frac{\alpha}{n+\alpha-1}p(\bx_i|\bphi^0)
\end{equation}
for the creation of a new basis.

\subsubsection{Using the predictive posterior distribution}
\label{section_imm_clustering}

Similar way to the approach used in Sec.~\ref{posterior_predictive_distribution}, we can directly predict the new outcomes in the case of an infinite mixture as a function of already observed data.

The probabilities of a new assignation will be
\begin{equation}
    p_\infty(z_{lk}=1|\set{\bz}_{\neg l},\set{\bphi},\alpha) \propto \frac{n_{\neg l}}{n+\alpha-1}p(\bx_i|\set{\bx}_k)
\end{equation}
for a basis with a populated sample and
\begin{equation}
    p_\infty(z_{l,K+1}=1|\set{\bz}_{\neg l},\set{\bphi},\alpha) \propto \frac{\alpha}{n+\alpha-1}p(\bx_i|\bphi^0)
\end{equation}
for a new basis.

\subsection{Modifications of the convolution distribution}
\label{modifications_convolution_distribution}

Introducing indicator variables as indicated in \ref{fmm_likelihood}, the convolution of two multivariate mixture distributions like the one described by \eqref{eq:pc} takes the form 
\begin{equation}
    p(\set{\bx},\set{\bz}|\set{\bphin},\bwn,\set{\bphitar},\bwtar)=\prod_i^N\prod_j^{\noise{K}}\prod_k^{\target{K}} \N(\bx_i|\bmun_j+\bmutar_k,\bSigman_j+\bSigmatar_k)^{z_{ijk}}(\wn_j\wtar_k)^{z_{ijk}}
\end{equation}
where the indicator variable is now 1 if the sample $i$ belongs to the noise base $j$ and target base $k$. Considering the approximation described in \ref{Posterior_approximation}, the only sampling parameters that we have to go over will be $\set{\bphitar}$, the weights $\bwtar$ and the indicator variables $\bz$. The main challenge is that there is no close form to group all the terms involving $\Sigmatar$ in single effective distributions as the ones derived in \ref{multivariate_normal_distribution}. In order to get a tractable expression, we would like to transform the convoluted covariances in such a way that the following expression follows:
\begin{equation}
    \bm{M}(\bSigman_j+\bSigmatar_k)^{-1}=(\bm{M}+\bm{A}){\bSigmatar_k}^{-1}
\end{equation}
for given matrices $\bm{A}$, $\bm{M}$. Isolating, $\bm{A}$ we obtain that
\begin{equation}
    \bm{A} = -\bm{M}(\bSigman_j+\bSigmatar_k)^{-1}\bSigman_j
\end{equation}

Now, in most practical cases we can consider that the convoluted covariance will be close to the expected covariance matrix, and we can approximate the expression above by
\begin{equation}
    \bm{A} \sim -\bm{M}\hat{\bSigmaInv}_{jk}\bSigman_j,
\end{equation}
where the effective expected covariance is
\begin{equation}
    \hat{\bSigma}_{jk} = n_{jk}\bs{jk}+\kappa_0(\mbx_{jk}-\bmu_0)(\mbx_{jk}-\bmu_0)^T+ \bSigma_0
\end{equation}
All the results derived in the preceding sections take into account this approximation:
\begin{align}
    \bmu &\rightarrow \bmutar+\bmun\\
    \bSigma &\rightarrow \bSigmatar+\bSigman\\
    \bSigmaInvtar &\rightarrow (\mathbb{I}-\hat{\bSigmaInv}_{jk}\bSigman_j)\bSigmaInvtar\\
    z_{ij} &\rightarrow z_{ijk}
\end{align}

The weights of the target distribution (\ref{conditional_distribution_weights}) will be computed using the sum over the samples for all the noise samples
\begin{align}
    \target{n}_k = \sum{ij}z_{ijk}
\end{align}
Sampling from the mean and covariance parameters in the distribution needs, however, a closer look. 

\subsubsection{Prior distribution}

In order for the prior to be conjugated, we have to scale the prior distribution in terms of the prior of the noise distribution.
\begin{align}
&p(\bmu|\bSigmaeff) =  \N(\bmu;\bmu_0,\kappa_0\bSigmaeff)\\
&p(\bSigmaeff) =  \InvW(\bSigmaeff;\bSigma_0,\bSigman,\nu_0)   
\label{eq:mn_conjugate_prior_conv}
\end{align}
where the effective covariance is
\begin{equation}
    \bSigmaeff=\bSigmatar+\bSigman
\end{equation}

\subsubsection{Conditional distribution: covariance}

If we focus the analysis in a single set of target parameters $\bphi_k$, the posterior probability of the variance and the mean has the form
\begin{equation}
p(\bphitar_k|\set{\bphin},\set{\bx},\set{\bz})\propto \prod_i^N\prod_j^{K_\xi} \N(\bx_i|\bmun_j+\bmutar_k,\bSigman_j+\bSigmatar_k)^{z_{ijk}} p(\bmu|\bSigman_j+\bSigmatar_k)p(\bSigman_j+\bSigmatar_k)
\end{equation}

In contrast with the case without convolution, the normal distributions cannot be grouped together in a simple distribution that can be sampled by standard procedures. We can group the data coming from a specific dataset,
\begin{equation}
    p(\bphitar_k|\set{\bphin},\set{\bx},\set{\bz})\propto
    \det{(\bSigman_j+\bSigmatar_k)}^{-(n_{jk}+\nu_0+1)/2}\Normal{\bSigmaeff_{jk}(\bSigman_j+\bSigmatar_k)^{-1}}
\end{equation}
where the variance is
\begin{equation}
    \bSigmaeff_{jk} = n_{jk}\bs{jk}+n_{jk}(\mbx_{jk}-\bmun_{j}-\bmutar_{k})(\mbx-\bmun_{j}-\bmutar_{k})^T+\kappa_0(\bmutar_{k}-\bmu_0)(\bmutar_{k}-\bmu_0)^T+ \bSigma_0
\end{equation}

Within the proposed approximation, we can now group all the different distributions:
\begin{equation}
    \bSigmaeff_{k} = \sum_j \bSigmaeff_{jk}(\mathbb{I}-\hat{\bSigmaInv}_{jk}\bSigman_{j})
\end{equation}
and we can sample new covariance matrices for the target with the inverse Wishart distribution.
\begin{equation}
    \bSigmatar_{k}\sim \InvW(\bSigmaeff_{k},n_{k}+\nu_0+1)
\end{equation}

\subsubsection{Conditional distribution: mean}

The mean distribution can be easily computed, regrouping all the terms in the following summary statistics:
\begin{align*}
    \bSigmaInveff_k &= \sum_{j}(n_{jk}+\kappa_0)(\bSigman_{j}+\bSigmatar_{k})^{-1}\\
    \bmueff_k &= \bSigmaeff_j\left(\sum_{j}(\bSigman_{j}+\bSigmatar_{k})^{-1}(n_{jk}\mbx_{jk}+\kappa_0\bmu_0)\right),\\
\end{align*}
and sampling from a multivariate normal,
\begin{equation}
    \bmutar_{k}\sim \N(\bmueff_k,\bSigmaInveff_k)
\end{equation}

\section{Hyperparameter selection}

Our approach the has five hyperparameters:
\begin{itemize}
    \item $\alpha$: Determines how close we are from a uniform distribution of weights (finite mixtures), or the potential of generating a new basis (infinite mixture).
    \item $\bmu_0$: The center of the prior normal distribution $\mu_0$.    
    \item $\kappa_0$: The confidence we have of being close to $\mu_0$.
    \item $\bSigma_0$: The covariance of the prior normal distribution.
    \item $\nu_0$: The confidence we have of being close to $\sigma_0$.    
\end{itemize}
Flow cytometry applications have in general have large datasets, making the approach quite insensitive to the choice of $\alpha$, $\kappa_0$ and $\nu_0$, as the statistics will dominate the model. The choice of mean and covariance matrix, however, depend on the data distribution. Appropriate estimators of these parameters are the mean and the covariance matrix of the whole dataset. This imposes a soft-informative prior over the region where the density should lie. In cases where the distribution has fat tails, the density prior can be very flat because of the outliers, making it difficult for the algorithm to find the correct distribution as the prior spreads over the basis components. In these cases, narrower prior covariance matrices can help to fit the model correctly.

\section{Gibbs sampling algorithms}

In this section, we describe the algorithms to sample the noise distribution according to the approximation described in \ref{Posterior_approximation}. 

\subsection{Efficient updating of summary statistics}
\label{efficient_updating}

Once we have computed the mean or the variance over a set of samples $\set{\bx}$ and we add or remove a single sample, it is possible to update the statistic without having to recompute the metric fully over the new set, which in general will be more time-consuming as we will have to go over a sum over all the size of the set.

If we remove a sample:
\begin{align}
    n_j^{rem} &= n_j - 1\\
    \mbx_j^{rem} &= \frac{n_j\mbx_j-\bx_i}{n_j^{rem}}\\
    {\bs{j}}^{rem} &= \frac{1}{n_j^{rem}}(n_j\bs{j} +n_j\mbx_j\mbxT_j-\bx_i\bxT_i)-\mbx_j^{rem}\mbxT_{j,rem}
\end{align}

On the other hand, if we add a sample:
\begin{align}
    n_j^{add} &= n_j + 1\\
    \mbx_j^{add} &= \frac{n_j\mbx_j+\bx_i}{n_j^{add}}\\
    {\bs{j}}^{add} &= \frac{1}{n_j^{add}}(n_j\bs{j} +n_j\mbx_j\mbxT_j+\bx_i\bxT_i)-\mbx_j^{new}\mbxT_{j,add}
\end{align}

\subsection{Finite normal mixture distribution}

The procedure in this case is as follows:
\begin{itemize}
    \item[0] Initialize the parameters.
    \begin{itemize}
        \item[0.1] Initialize the indicator variables $\set{\bz}$: Assign to each cell one of the $K$ normal distributions using any initialization procedure (random assignment, k-means...).
        \item[0.2] Compute the statistics of each base:
        \begin{align*}
            n_j = \sum_i z_{ij}\hspace{1cm}
            \mbx_j = \frac{1}{n_j}\sum_i z_{ij}\bx_i\hspace{1cm}
            \bs{j} = \frac{1}{n_j}\sum_i z_{ij}\bx_i\bxT_i-\mbx_j\mbxT_j
        \end{align*}
        \item[0.3] Initialize parameters of each base:
        \begin{align*}
            \bmu_j = \mbx_j\hspace{1cm}
            \bSigma_j = \bs{j}\hspace{1cm}
            \w_j = \frac{1}{n}\sum_i z_{ij}
        \end{align*}
        where $n=\sum_jn_j$ is the size of the dataset.
    \end{itemize}
    \item[1] Sampling. For $N$ iterations do
    \begin{itemize}
        \item[1.1] Sample indicator variables as in \ref{conditional_distribution_indicator_variables}. For each sample $i$:
        \begin{itemize}
            \item[1.1.1]Compute weights: $w_j=\w_jp(\bx_i|\bphi_j)$.
            \item[1.1.2]Sample new indicator: $\bz \sim \Multinomial(1,\bm{w})$.
        \end{itemize}
        \item[1.2] Sample weights as from \ref{conditional_distribution_weights}. For each base $j\in{1,...K}$ do
        \begin{itemize}
            \item[1.2.1]Compute $\bneff$: $\neff_j = \sum_iz_{ij}+\alpha/K$
            \item[1.2.2]Sample new weights: $\bw \sim \Dirichlet(\bneff)$
        \end{itemize}        
        \item[1.3] Sample variances and mean. For each base $j\in{1,...K}$ do
        \begin{itemize}
            \item[1.3.1]Compute summary statistics:
            \begin{align*}
                n_j = \sum_i z_{ij}\hspace{1cm}
                \mbx_j = \frac{1}{n_j}\sum_i z_{ij}\bx_i \hspace{1cm}
                \bs{j} = \frac{1}{n_j}\sum_i z_{ij}\bx_i\bxT_i-\mbx_j\mbxT_j
            \end{align*}
            \item[1.3.2]Sample new covariance \ref{conditional_distribution_covariance}:
            \begin{align*}
                \neff_j &= n_j+\nu_0+1\\
                \bSigmaeff_j &= n_j\bs{j} +n_j(\mbx-\bmu_j)(\mbx-\bmu_j)^T+\kappa_0(\bmu_j-\bmu_0)(\bmu_j-\bmu_0)^T+ \bSigma_0\\
                \bSigma_{j} &\sim \InvW(\neff_j,\bSigmaeff_j)
            \end{align*}
            \item[1.3.2]Sample new mean \ref{conditional_distribution_mean}:
            \begin{align*}
                \bmueff &= \frac{n_j\mbx_j+\kappa_0\bmu_0}{n_j+\kappa_0}\\
                \bSigmaInveff &= (n_j+\kappa_0)\bSigmaInv_j\\
                \bmu_{j} &\sim \N(\bmueff,\bSigmaeff)
            \end{align*}
        \end{itemize}        
    \end{itemize}
\end{itemize}

\subsection{Infinite normal mixture distribution}

The approach in this case has the following steps:
\begin{itemize}
    \item[0] Initialize the parameters.
    \begin{itemize}
        \item[0.1] Initialize the indicator variables $\set{\bz}$: Assign to each cell one of the $K$ normal distributions using any initialization procedure (random assignment, k-means...).
        \item[0.2] Compute the statistics of each base:
        \begin{align*}
            n_j = \sum_i z_{ij}\hspace{1cm}
            \mbx_j = \frac{1}{n_j}\sum_i z_{ij}\bx_i\hspace{1cm}
            \bs{j} = \frac{1}{n_j}\sum_i z_{ij}\bx_i\bxT_i-\mbx_j\mbxT_j
        \end{align*}
        \item[0.3] Initialize parameters of each base:
        \begin{align*}
            \bmu_j = \mbx_j\hspace{1cm}
            \bSigma_j = \bs{j}\hspace{1cm}
            \w_j = \frac{1}{n}\sum_i z_{ij}
        \end{align*}
        where $n=\sum_jn_j$ is the size of the dataset.
    \end{itemize}
    \item[1] Sampling. For $N$ iterations, do
    \begin{itemize}
        \item[1.1] Reassign samples. For each sample $i\in\set{1,...,n}$
        \begin{itemize}
            \item[1.1.1] Remove sample $i$. Consider sample $z_{ij} = 1$.
            
            If $n_j=1$ (the only sample assigned to that base distribution), remove the distribution from the active basis.
            
            Otherwise recompute the summary statistics for base distribution $j$ removing one sample as described in \ref{efficient_updating}.

            \item[1.1.2] Compute weights (not-normalized) for reassigning the sample as described in sections \ref{section_imm_complete} and \ref{posterior_predictive_distribution}.
            
            For the active bases $j \in \set{1,...K}$ . 
            \begin{align*}
                &m^0_j = \frac{n_j+\kappa_0}{n_j+\kappa_0+1}\hspace{1cm}
                m^1_j = \frac{\frac{n_j\kappa_0}{n_j+\kappa_0}+n_j\kappa_0}{n_j+\kappa_0+1}\\
                &\bmueff^y_j = \frac{(n_j\mbx+\kappa_0\bmu_0)}{n_j+\kappa_0}\hspace{1cm}
                \bSigmaeff^y_j = \frac{1}{m^0}\left(m^1(\bmu_0-\mbx)(\bmu_0-\mbx)^T +n_j\bs{x} + \bSigma_0\right)\\
                &w_{j} = \frac{n_j}{n+\alpha-1}\TStudent{\nueff}(\by;\bmueff^y,\bSigmaeff^y/\nueff)
            \end{align*}
            and for creating a new basis,
            $$w_{K+1} = \frac{\alpha}{n+\alpha-1}\N(\bx_i|\bmu^y_0,\bSigma^y_0)$$
            \item[1.1.3]Sample new indicator: $\bz_{i}^{new} \sim \Multinomial(1,\bm{w})$.
            \item[1.1.4] Update statistics. 
            
            If $z_{iK+1}=1$ create a new basis distribution and assign new statistics to it,
            \begin{align*}
                n_{K+1} = 1\hspace{1cm}
                \mbx_{K+1} = \bx_i\hspace{1cm}
                \bs{K+1} = 0
            \end{align*}
            else, update the statistics adding a term as described in  \ref{efficient_updating}.
        \end{itemize}
        \item[1.2] Sample basis parameters. For each basis $j \in \set{1,...,K}$:
        \begin{itemize}
            \item[1.2.1] Sample the weights \ref{conditional_distribution_weights}:
                \begin{align*}
                    \neff_j=n_j+\alpha\hspace{1cm}
                    \bw \sim \Dirichlet(\bneff)\\
                \end{align*}   
            \item[1.2.2] Sample the covariance matrix \ref{marginal_distribution_variance}:
                \begin{align*}
                    \neff_j = n_j+\nu_0+1\hspace{1cm}
                    \bSigmaeff_j = n\bs{j}+\frac{\kappa_0n_j}{\kappa_0+n_j}(\mbx_j-\bmu_0)(\mbx_j-\bmu_0)^T +\bSigma_0
                \end{align*}
                \begin{align*}
                    \bSigma_j\sim\InvW(\bSigmaeff_j,\neff_j)
                \end{align*}
            \item[1.2.3] Sample the mean \ref{conditional_distribution_mean}:
                \begin{align*}
                    \bmueff_j = \frac{n_j\mbx_j+\kappa_0\bmu_0}{n_j+\kappa_0}\hspace{1cm}
                    \bSigmaeff_j = \bSigma_j/(n_j+\kappa_0)
                \end{align*}
                \begin{align*}
                    \bmu_{j} \sim \N(\bmueff_j,\bSigmaeff_j)
                \end{align*}
        \end{itemize}
    \end{itemize}
\end{itemize}

\subsection{Modifications to the convoluted distribution}

The preceding algorithms are the sampling algorithms for non-convoluted finite and infinite mixture sampling. The convoluted cases are exactly the same, but changing the equations with the modifications described in section \ref{modifications_convolution_distribution}. In addition to this, it is necessary to add a step in the sampling loop to sample new parameters from the already fitted noise distribution $\set{\bphin_j}_j$.

\section{Efficiency assessment}

So far, the efficiency of deconvolution methods has been assessed using quantifiers applicable to point estimates, which are the ones proposed in the literature to date.
To compare the efficiency of our model with previous methods that do not obey the positivity nor the normalization conditions, we can use the Mean Integrated Squared Estimation (MISE) measure:
\begin{equation}
\text{MISE} = \int_{-\infty}^\infty(p_{\rm true}(\bx)-p_{\rm est}(\bx))^2dx,
\end{equation}
where $p_{\rm true}$ is the real target distribution and $p_{\rm est}$ is the point estimate of the deconvolved distribution.
This is the traditional measure of convergence, but it is hard to interpret as it only has a lower bound.
To address this issue, we introduce the mean integrated overlap (MIO): 
\begin{equation}
\text{MIO} = 1-\frac{1}{2}\int_{-\infty}^\infty|p_{true}(\bx)-p_{est}(\bx)|dx
\end{equation}
This measure has the property that it is bounded in the interval $[0,1]$ if the true and the estimated distributions are normalized, with 0 corresponding to the case of no overlap between distributions, and 1 to the case of complete overlap.
Values below zero can be obtained if the estimated distribution does not follow the positivity requirement nor the normalization condition, as it is the case in FFT-based deconvolutions.

\bibliographystyle{vancouver}
\bibliography{deconv}
\addcontentsline{toc}{section}{References}

\newpage
\section*{Supplementary figures}
\addcontentsline{toc}{section}{Supplementary figures}

\begin{figure}[htbp]
\centering
\includegraphics[width=0.95\textwidth]{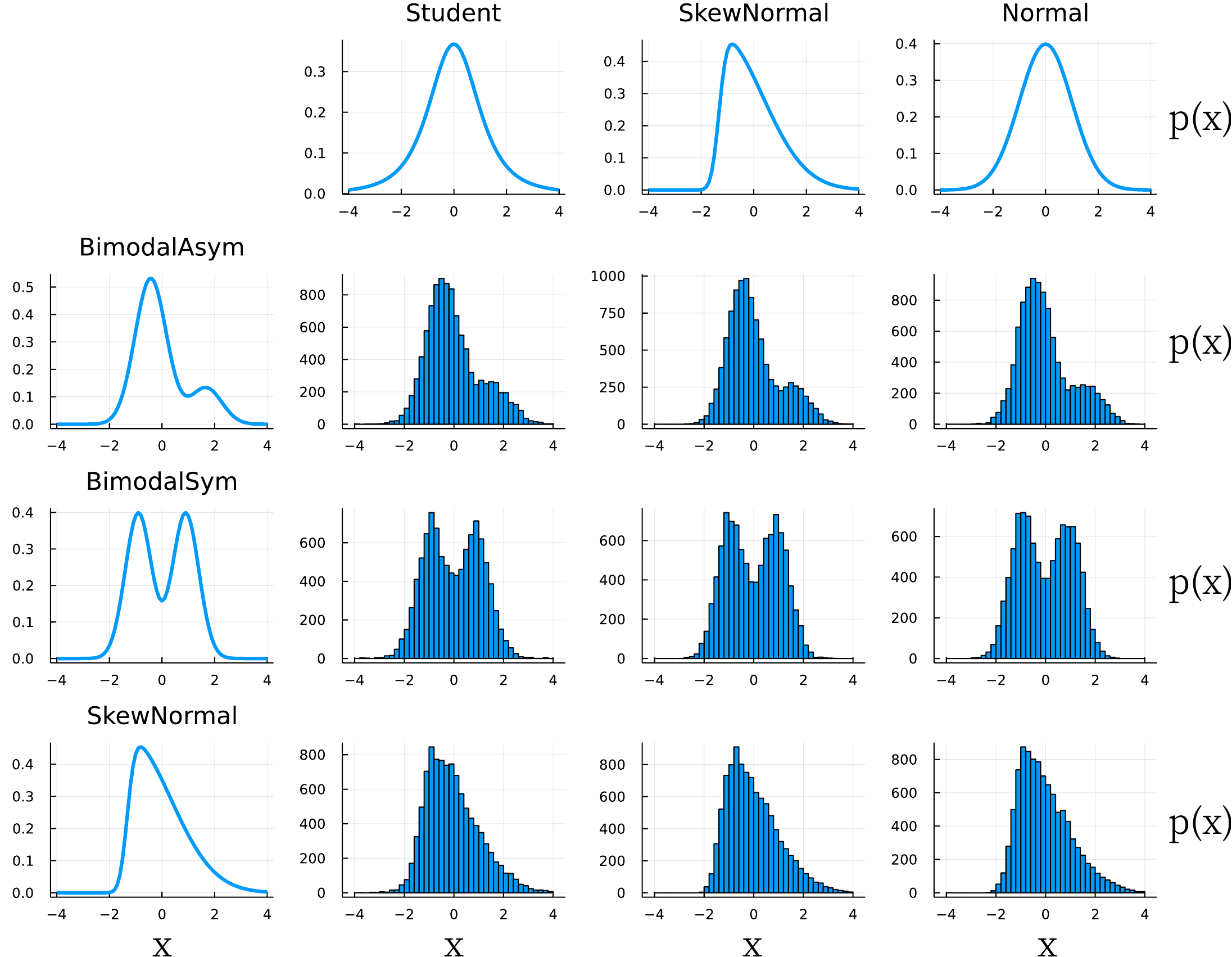}
\caption{Synthetic target distributions (left) and noise distributions (top) and the resulting convolutions for a SNR=2.}
\end{figure}

\newpage
\begin{figure}[htp]
\centering
\includegraphics[width=0.95\textwidth]{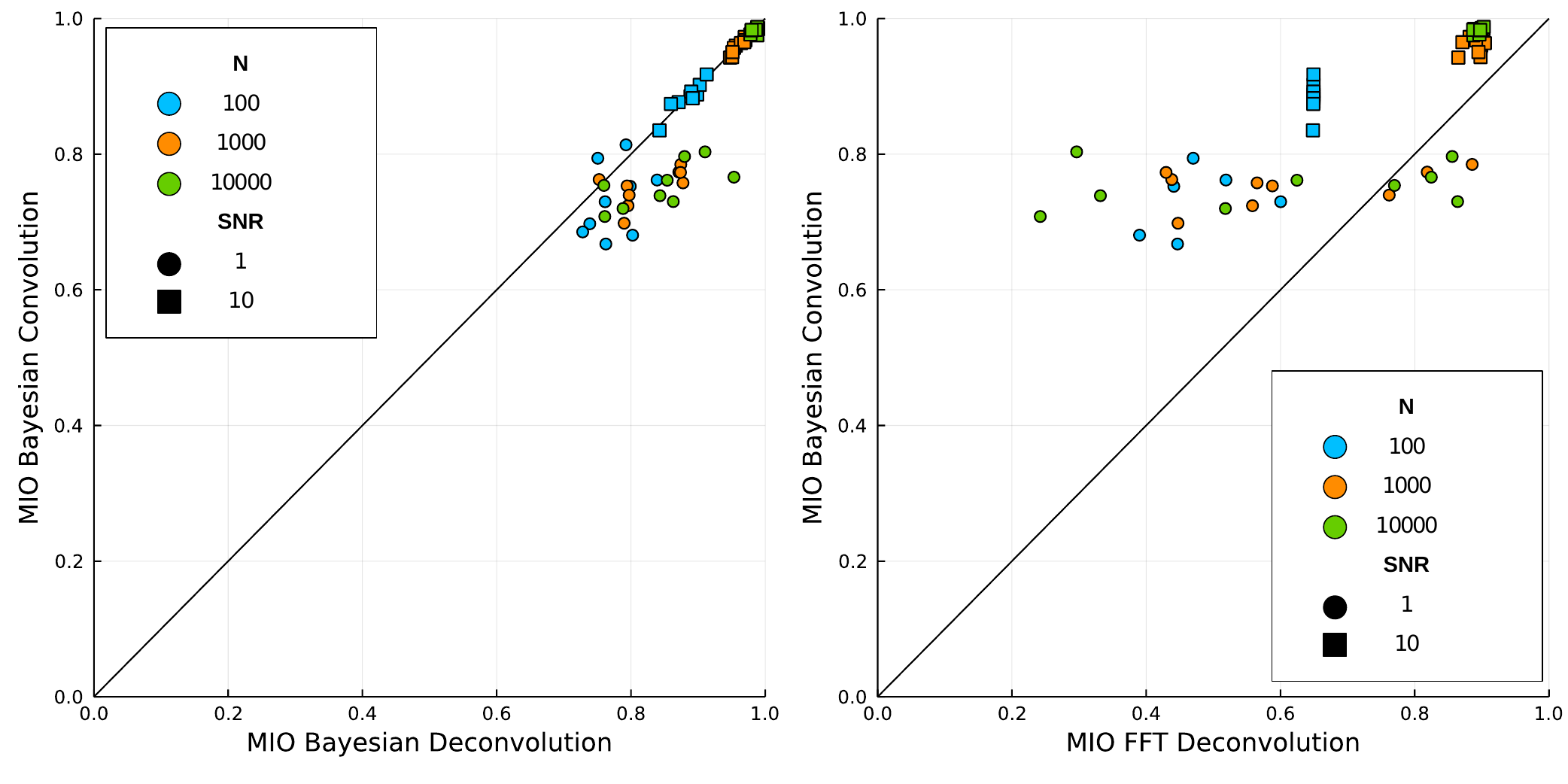}
\caption{Comparison between the deconvolved and ground-truth target distributions as expressed by the Mean Integrated Overlap (MIO) for the Bayesian (left) and FFT (right) methods (x axis), with the results of a null Bayesian model fitting directly to the convolved data (y axis), ignoring the noise.}
\end{figure}

\newpage
\begin{figure}[htbp]
\centering
\includegraphics[width=0.95\textwidth]{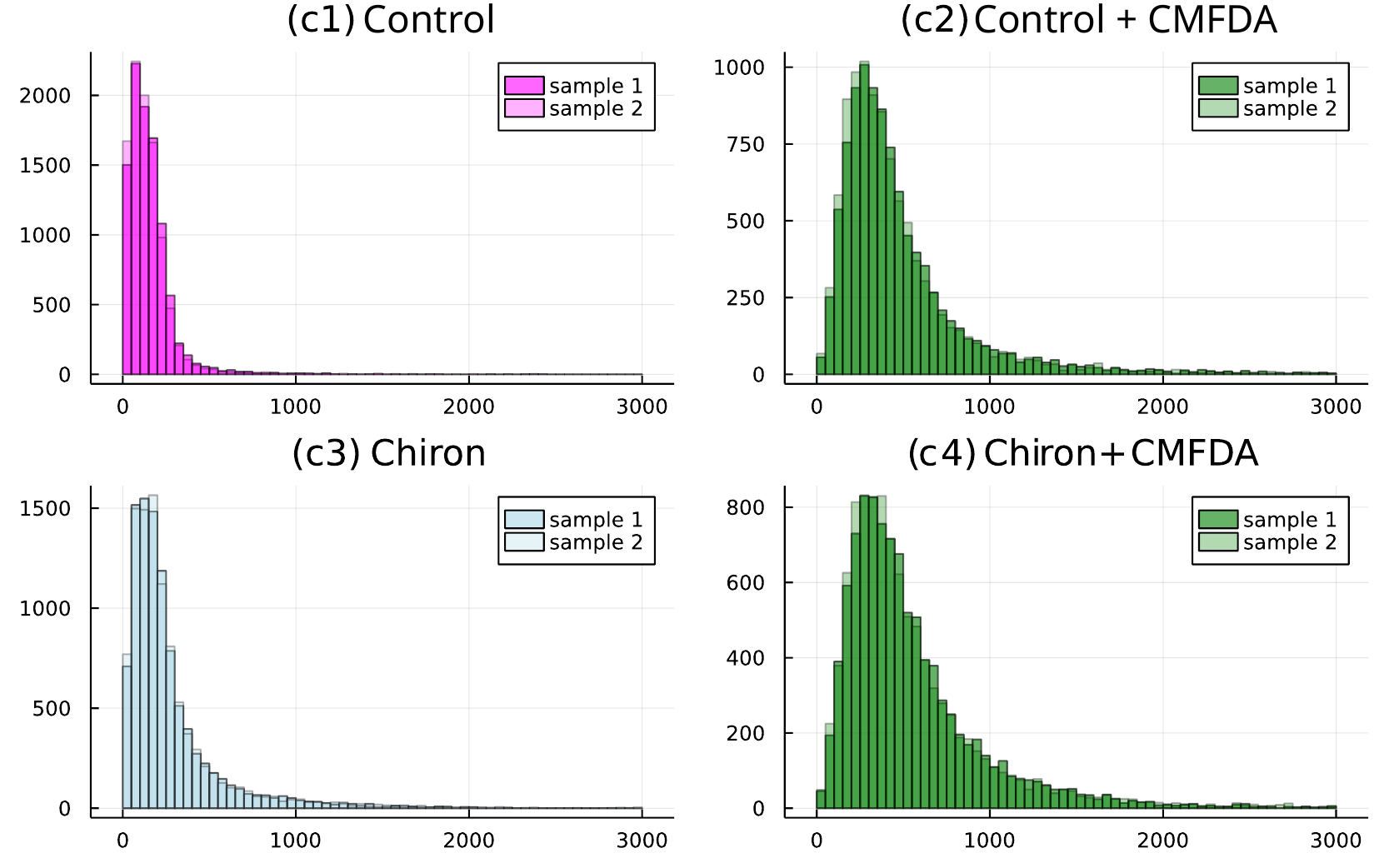}
\caption{Flow cytometry distributions obtained in the four experimental conditions discussed in Sec.~III.B of the main text.} 
\end{figure}

\newpage
\begin{figure}[htbp]
\centering
\includegraphics[width=0.95\textwidth]{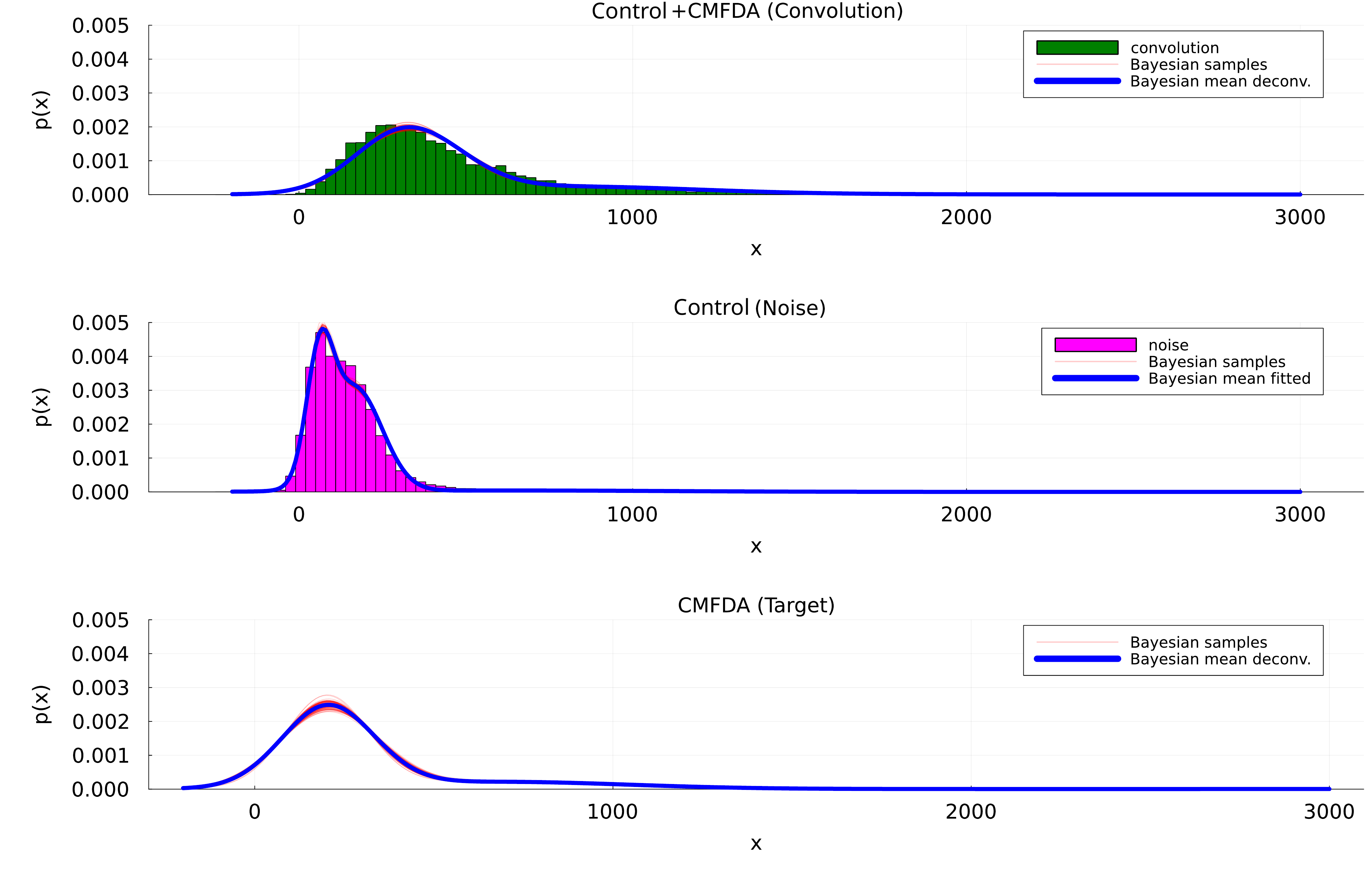}
\caption{Flow cytometry distributions corresponding to the data conditions c2 (top panel in green) and c1 (middle panel in magenta) discussed in Sec.~III.B. Overlayed on the distributions we show realizations of the Bayesian sampling process (red lines) for the three distributions (noise, convolution and target) obtained during the fitting.} 
\end{figure}

\newpage
\begin{figure}[h!]
\centering
\includegraphics[width=0.95\textwidth]{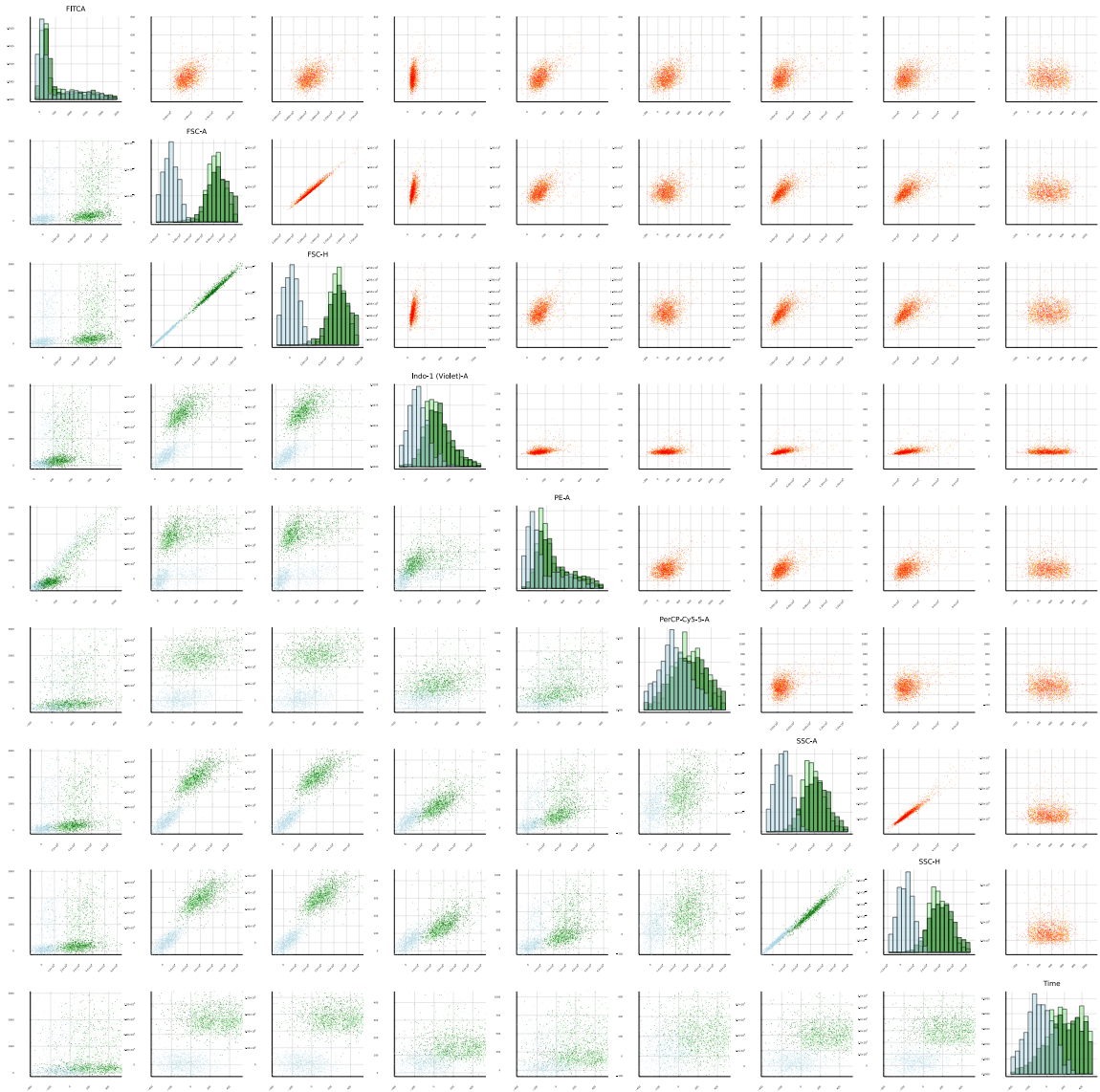}
\caption{Scatter plot sections of the multichannel dataset. The top right triangle shows the autofluorescence distribution (orange) and samples from the fitting process (red). In the low left triangle we show the convolution data (light green) and samples from fitted convolution (dark green) and the deconvolution (light blue). In the diagonal, we show the 1D distributions of the convolved and deconvolved results.}
\end{figure}